\pdfoutput=1
\documentclass[11pt,twoside,a4paper,cmspaper,final,collab]{cms-tdr}
\def\svnVersion{97f0408}\def\svnDate{2023/05/23}\def\cmsCernNoTag{CERN-EP-2022-117}\def\cmsCernDate{\today}\def\cmsMessage{Published in Physics Letters B as \href{http://dx.doi.org/10.1016/j.physletb.2022.137531}{\doi{10.1016/j.physletb.2022.137531}.}}
\begin{document}\cmsNoteHeader{HIG-20-010}

\newcommand{\muth}{\ensuremath{\PGt_\PGm\tauh}\xspace}
\newcommand{\eleth}{\ensuremath{\PGt_\Pe\tauh}\xspace}
\newcommand{\thth}{\ensuremath{\tauh\tauh}\xspace}
\newcommand{\bbtt}{\ensuremath{\PQb\PQb\PGt\PGt}\xspace}
\newcommand{\lambdahhh}{\ensuremath{\ensuremath{\lambda_{\PH\PH\PH}}\xspace}\xspace}
\newcommand{\HH}{\ensuremath{\PH\PH}\xspace}
\newcommand{\WW}{\ensuremath{\PW\PW}\xspace}
\newcommand{\WZ}{\ensuremath{\PW\PZ}\xspace}
\newcommand{\ZZ}{\ensuremath{\PZ\PZ}\xspace}
\newcommand{\ggF}{\ensuremath{\Pg\Pg\text{F}}\xspace}
\newcommand{\pp}{\ensuremath{\Pp\Pp}\xspace}
\newcommand{\ttH}{\ensuremath{\PQt\PAQt\PH}}

\newlength\cmsTabSkip\setlength{\cmsTabSkip}{1ex}
\ifthenelse{\boolean{cms@external}}{\providecommand{\cmsLeft}{upper\xspace}}{\providecommand{\cmsLeft}{left\xspace}}
\ifthenelse{\boolean{cms@external}}{\providecommand{\cmsRight}{lower\xspace}}{\providecommand{\cmsRight}{right\xspace}}

\cmsNoteHeader{HIG-20-010}
\title{Search for nonresonant Higgs boson pair production in final state with two bottom quarks and two tau leptons in proton-proton collisions at \texorpdfstring{$\sqrt{s} = 13\TeV$}{sqrt(s) = 13 TeV}}

\author*[inst1]{Francesco Brivio}

\date{\today}

\abstract{
   A search for the nonresonant production of Higgs boson pairs (\HH) via gluon-gluon and vector boson fusion processes in final states with two bottom quarks and two tau leptons is presented. The search uses data from proton-proton collisions at a center-of-mass energy of $\sqrt{s}=13\TeV$ recorded with the CMS detector at the LHC, corresponding to an integrated luminosity of 138\fbinv. Events in which at least one tau lepton decays hadronically are considered and multiple machine learning techniques are used to identify and extract the signal. The data are found to be consistent, within uncertainties, with the standard model (SM) predictions. Upper limits on the \HH production cross section are set to constrain the parameter space for anomalous Higgs boson couplings. The observed (expected) upper limit at 95\% confidence level corresponds to 3.3 (5.2) times the SM prediction for the inclusive \HH cross section and to 124 (154) times the SM prediction for the vector boson fusion \HH cross section. At 95\% confidence level, the Higgs field self-coupling is constrained to be within $-1.7$ and 8.7 times the SM expectation, and the coupling of two Higgs bosons to two vector bosons is constrained to be within $-0.4$ and 2.6 times the SM expectation.
}

\hypersetup{
pdfauthor={CMS Collaboration},
pdftitle={Search for nonresonant Higgs boson pair production in final states with two bottom quarks and two tau leptons in proton-proton collisions at sqrt(s) = 13 TeV},
pdfsubject={CMS},
pdfkeywords={CMS,  Higgs, HH, bbtautau, self-coupling, VBF}}

\maketitle

\section{Introduction}
\label{sec:intro}
The discovery of the Higgs boson (\PH) by the ATLAS and CMS Collaborations~\cite{Aad:2012tfa,Chatrchyan:2012ufa,Chatrchyan:2013lba} was a major step towards improving the understanding of the mechanism of electroweak symmetry breaking. With the value of the \PH{} mass ($m_\PH$) experimentally measured by the CMS Collaboration to be ${125.38 \pm 0.14\GeV}$~\cite{2020135425}, the structure of the Higgs scalar field potential and the strength of the \PH{} self-coupling are precisely predicted in the standard model (SM). While the measured properties are so far consistent with the SM~\cite{Khachatryan:2016vau}, determining the \PH{} self-coupling provides an independent test of the SM and is fundamental to probe experimentally the shape of the scalar potential that is at the base of the electroweak symmetry breaking mechanism.

\begin{figure*}[!htb]
    \centering
    \includegraphics[width=0.9\textwidth]{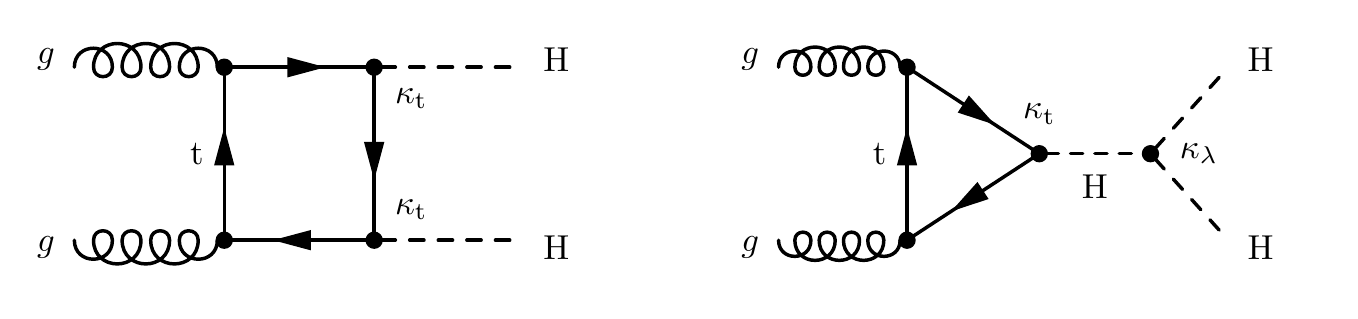}
    \caption{Feynman diagrams contributing to \HH production via gluon-gluon fusion in the SM at leading order.
    The different \PH{} interactions are labelled by the coupling modifiers $\kappa$.}
    \label{fig:Feyndiag_ggF}
\end{figure*}

The trilinear self-coupling of the Higgs boson (\lambdahhh) can be extracted from the measurement of the Higgs boson pair (\HH) production cross section. In the SM, the \HH production occurs for proton-proton (\pp) collisions at the LHC mainly via gluon-gluon fusion (\ggF), involving either couplings to a loop of virtual fermions, or the \lambdahhh coupling itself. The leading-order (LO) $\Pg\Pg\to\HH$ Feynman diagrams shown in Fig.~\ref{fig:Feyndiag_ggF} have approximately the same amplitude but interfere destructively, therefore reducing the production cross section. The SM prediction for the \HH{} production cross section, computed at the next-to-next-to-LO (NNLO) precision for a center-of-mass energy of $\sqrt{s}=13\TeV$, is $\sigma_{\ggF}^{\HH}=31.05^{+6\%}_{-23\%}\text{ (scale}+m_{\PQt}\text{)}\pm3\%\text{ (PDF}+\alpS\text{)}\unit{fb}$ for an $m_\PH$ of 125\GeV~\cite{Grazzini_2018,PhysRevD.103.056002}, where PDF is the parton distribution function, \alpS is the strong coupling constant and $m_{\PQt}$ is the mass of the top quark.

\begin{figure*}[!htb]
    \centering
    \includegraphics[width=0.9\textwidth]{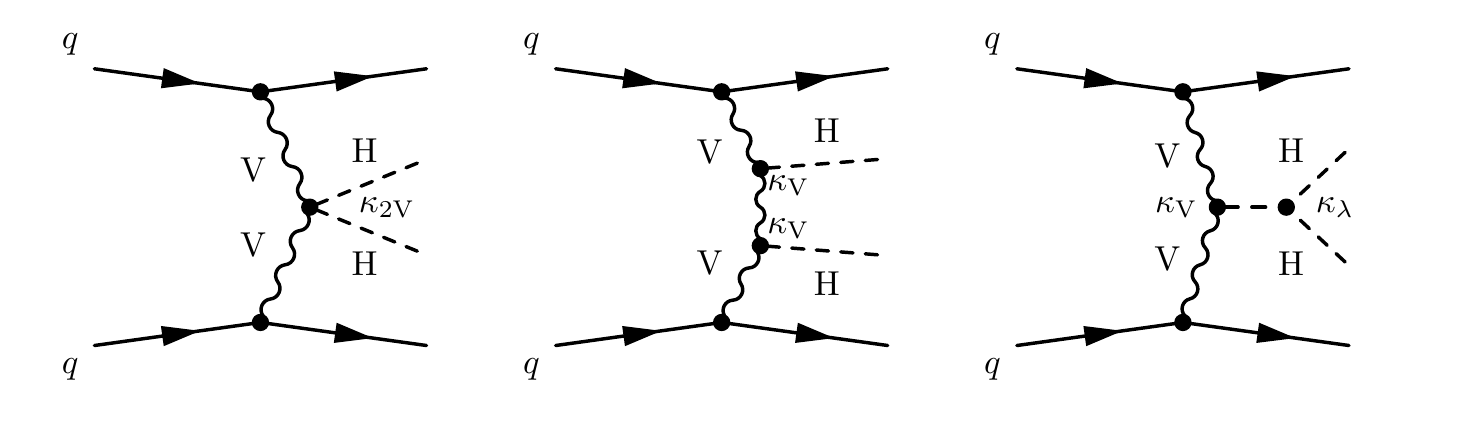}
    \caption{Feynman diagrams contributing to Higgs boson pair production via vector boson fusion in the SM at leading order.
    The different \PH{} interactions are labelled with the coupling modifiers $\kappa$.}
    \label{fig:Feyndiag_VBF}
\end{figure*}

Among the other \HH{} production modes, vector boson fusion (VBF) is of particular interest as it is the second-largest contribution to \HH production and
it gives access to the coupling between one or two \PH{} and two vector bosons. Despite its small production cross section,
$\sigma_{\text{VBF}}^{\HH}=1.726^{+0.03\%}_{-0.04\%}\text{ (scale)}\pm2.1\%\text{ (PDF}+\alpS\text{)}\unit{fb}$ at next-to-NNLO precision for $\sqrt{s}=13\TeV$~\cite{PhysRevD.98.114016},
the distinctive VBF topology provides a useful handle to identify signal events, thanks to its two forward jets being well separated in pseudorapidity ($\eta$) and with a large invariant mass. The LO Feynman diagrams for VBF \HH production are shown in Fig.~\ref{fig:Feyndiag_VBF}.

Beyond SM physics effects in the nonresonant case can appear via anomalous couplings of the \PH{} or via new particles that contribute to the virtual quantum loops in Fig.~\ref{fig:Feyndiag_ggF}. The experimental signature would be a modification of the \HH{} production cross section and of the event kinematical properties~\cite{DiMicco:2692014}.

The variation of \lambdahhh with respect to the SM prediction is parameterized by the coupling modifier $\kappa_{\lambda}=\lambdahhh/\lambdahhh^{\text{SM}}$,
which affects both the \ggF{} and the VBF production mechanisms. Similarly, the variations of the coupling of the \PH{} to the top quark are parameterized as $\kappa_{\text{t}}$, while those of the couplings involved in the VBF production are parameterized as $\kappa_{\PV}$ ($\text{\PV\PV}\PH$ coupling) and $\kappa_{2\PV}$ ($\text{VV}\HH$ coupling).

In this paper, we present a search for \HH{} production in \pp collisions at $\sqrt{s}=13\TeV$, investigating both the \ggF{} and the VBF production mechanisms. The final state where one \PH{} decays to $\PQb\PAQb$ and the other decays to $\PGt^+\PGt^-$ (for simplicity $\HH\to\bbtt$), characterized by a decay branching fraction of 7.3\% for $m_{\PH{}}=125\GeV$, is studied. The sizable branching fraction, combined with the precise tau lepton identification algorithms developed within the CMS Collaboration (\textsc{DeepTau}~\cite{CMS:2022prd}), makes this final state one of the most sensitive to \HH{} production. We consider three decay modes of the $\PGt\PGt$ system: \muth, \eleth, and \thth, where $\PGt_\PGm$, $\PGt_\Pe$, and \tauh correspond, respectively, to the decay of a tau lepton into a muon, an electron, or hadrons, with the associated neutrinos. These account for 87.6\% of the full $\PGt\PGt$ decay modes. Quantum chromodynamics (QCD) multijet and $\PZ/\PGg^*\to\Pell\Pell$ events represent the dominant background in the fully hadronic channel, while in the semileptonic channels \ttbar events represent the largest contamination. The data used in this analysis correspond to an integrated luminosity of 138\fbinv collected during 2016--2018 by the CMS experiment.

The search described in this paper improves on the previous $\HH\to\bbtt$ results from the CMS and ATLAS Collaborations~\cite{Sirunyan:2273798,PhysRevLett.121.191801} by making use of a larger data sample, an improved trigger strategy, and the introduction of several newly developed neural networks that identify the \PQb{} jets from the \PH{} decay, categorize the events, and perform signal extraction. Moreover, this analysis builds up on CMS-wide innovations such as the new \textsc{DeepTau}~\cite{CMS:2022prd} and \textsc{DeepJet}~\cite{Bols_2020} algorithms. All these factors lead to achieving particularly stringent results on the \HH production cross sections.

Tabulated results are provided in the HEPData record for this analysis~\cite{hepdata}.

\section{The CMS detector}
\label{sec:detector}
The CMS detector is a multipurpose apparatus optimized to study high-transverse-momentum (\pt) physics processes in \pp collisions. The central feature of the CMS detector is a superconducting solenoid, providing a magnetic field of 3.8\unit{T}. Within the solenoid volume are a silicon pixel and strip tracker that cover a region of $\abs{\eta}<2.5$, a lead tungstate crystal electromagnetic calorimeter (ECAL), and a brass and scintillator hadron calorimeter (HCAL), each composed of a barrel and two endcap sections that cover $\abs{\eta}<3$. Forward calorimeters extend the $\eta$ coverage, provided by the barrel and endcap detectors, to $\abs{\eta}<5$. Muons are detected in gas-ionization chambers embedded in the steel flux-return yoke outside the solenoid and covering an angle $\abs{\eta}<2.4$. Events of interest are selected using a two-tiered trigger system. The first level (L1), composed of custom hardware processors, uses information from the calorimeters and muon detectors to select events at a rate of around 100\unit{kHz} within a fixed latency of about 4\mus~\cite{Sirunyan:2020zal}. The second level, known as the high-level trigger, consists of a farm of processors running a version of the full event reconstruction software optimized for fast processing, and reduces the event rate to around 1\unit{kHz} before data storage~\cite{Khachatryan:2016bia}.

A more detailed description of the CMS detector, together with a definition of the coordinate system used and the relevant kinematic variables, can be found in Ref.~\cite{Chatrchyan:2008zzk}.

\section{Data and simulated samples}
\label{sec:datasets}
The data used in this analysis correspond to an integrated luminosity of 36.3, 41.5, and 59.7\fbinv, for the 2016, 2017, and 2018 data-taking periods, respectively~\cite{CMS-LUM-17-003,CMS-PAS-LUM-17-004,CMS-PAS-LUM-18-002}.

Simulated samples of nonresonant \HH production via \ggF{} are generated at LO precision with \MGvATNLO v2.4.2~\cite{Alwall_2014} and at next-to-LO (NLO) precision with \POWHEG v2.0~\cite{Frixione_2007,Re_2011,Campbell_2015}. The former samples are used for training the machine learning algorithms implemented in this analysis, while the latter are used for the signal extraction: in order to model any arbitrary $\kappa_{\lambda}$ value, the expected distributions of signal events at three support points ($\kappa_{\lambda} = \{1.00, 2.45, 5.00\}$) are scaled by functions of the target value of $\kappa_{\lambda}$, defined according to the known dependence of the \ggF{} \HH cross section~\cite{Heinrich:2019bkc}, summed together, and then normalized to the corresponding NNLO cross section~\cite{Grazzini_2018}. Simulated samples of nonresonant \HH production via VBF are generated at LO precision with \MGvATNLO v2.3.2.2 (v2.6.0) for 2016 (2017--2018) and a similar procedure for modelling any $\kappa_{2\PV}$ value is applied as in the \ggF{} case.

{\tolerance=1000 Backgrounds arising from $\PZ/\PGg^*\to\Pell\Pell$ production and \PW boson production in association with jets are simulated with \MGvATNLO at LO with MLM merging~\cite{Alwall_2007}; the single top quark, single \PH, and \ttbar backgrounds are simulated at NLO precision with \POWHEG v2.0; backgrounds from diboson (\WW, \ZZ, and \WZ) and triboson (\PZ{}\PZ{}\PZ{}, \PZ{}\PZ{}\PW{}, \PZ{}\PW{}\PW{}, and \PW{}\PW{}\PW{}) production are simulated at NLO precision with \MGvATNLO, both with and without FxFx merging~\cite{Frederix_2012}, and \POWHEG, depending on the process; production of \ttbar pairs in association with a single boson or a pair of vectors bosons are simulated with \MGvATNLO, both with and without FxFx merging, depending on the process. For the simulation of backgrounds using \MGvATNLO, different versions of the generator are used (2.2.2, 2.3.2.2, 2.3.3, and 2.4.2), depending on the process and year. LO samples use the MLM merging allowing up to four extra partons in the matrix element, while for NLO samples generated with the FxFx merging only one extra parton is allowed.\par}

For both signal and background events, multiparton, parton shower, and hadronization effects are simulated with \PYTHIA v8.226 for 2016 and \PYTHIA v8.230 for 2017--2018~\cite{Sj_strand_2015}. For 2016, the CUETP8M1~\cite{Khachatryan_2016} tune is used for all the samples, except for the single top and \ttbar samples, which rely on the CP5~\cite{Sirunyan_2020} tune; for 2017--2018, the CP5 tune is used throughout all samples. The utilized parton distributions functions are the NNPDF 3.0~\cite{NNPDF_3_0} set for 2016 data sets and the NNPDF 3.1~\cite{NNPDF_3_1} set for 2017--2018 data sets. For all simulated samples, the CMS detector response is modeled with \GEANTfour~\cite{AGOSTINELLI2003250}. Additional collisions within the same or nearby bunch crossings (pileup) are simulated and added to the hard-scattering process for all samples. Events are then reweighted to match the pileup profile observed in data.

\section{Event reconstruction and selection}
\label{sec:objects}
The event reconstruction is based on the particle-flow (PF) algorithm~\cite{CMS:2017yfk}, which aims to reconstruct and identify each individual particle (PF candidate) in an event, with an optimized combination of information from the various elements of the CMS detector. PF candidates are classified as electrons, muons, photons, and charged or neutral hadrons. An important portion of the \bbtt final state is reconstructed and identified through the combination of PF objects, such as the jets originating from \PQb{} quarks, the \tauh candidates, and the missing transverse momentum vector \ptvecmiss, computed as the negative vector \pt sum of all the PF candidates in an event. The magnitude of the \ptvecmiss vector is denoted as \ptmiss~\cite{Sirunyan:2019kia}. The primary vertex (PV) is taken to be the vertex corresponding to the hardest scattering in the event, evaluated using tracking information alone, as described in Section 9.4.1 of Ref.~\cite{CMS-TDR-15-02}.

The electron momentum is estimated by combining the energy measurement in the ECAL with the momentum measurement in the tracker. The momentum resolution for electrons with $\pt \approx45\GeV$ from $\PZ\to\Pe\Pe$ decays ranges from 1.6 to 5.0\%. It is generally better in the barrel region than in the endcaps, and also depends on the bremsstrahlung energy emitted by the electron as it traverses the material in front of the ECAL~\cite{CMS:2020uim,CMS-DP-2020-021}. Electron identification (ID) is based on a boosted decision tree (BDT) that combines purely-ECAL, purely-tracker, and ECAL-plus-tracker observables, including the isolation of the electron. The ``tight'' working point of this discriminator, characterized by an 80\% efficiency, is employed in the analysis~\cite{CMS:2020uim}.

Muons are measured in the range $\abs{\eta}<2.4$, with detection planes made using three technologies: drift tubes, cathode strip chambers, and resistive plate chambers. The single muon trigger efficiency exceeds 90\% over the full $\eta$ range, and the efficiency to reconstruct and identify muons is greater than 96\%. Matching muons to tracks measured in the silicon tracker results in a relative \pt resolution, for muons with \pt up to 100\GeV, of 1\% in the barrel and 3\% in the endcaps. The \pt resolution in the barrel is better than 7\% for muons with \pt up to 1\TeV~\cite{CMS:2018rym}. The isolation of muon candidates ($\mathrm{I}_{\text{rel}}^{\ell}$) is defined as the \pt sum of PF candidates inside a $\Delta R=\sqrt{\smash[b]{(\Delta\eta)^2+(\Delta\phi)^2}}<0.4$ cone around the muon relative to the $\pt^{\ell}$ of the muon (where $\phi$ is the azimuthal angle in radians):
\begin{equation}
\frac{\sum \pt^\text{charged} + \max \left[ 0, \sum \pt^\text{neutral-had} + \sum \pt^{\gamma} - \frac{1}{2} \sum \pt^\mathrm{PU} \right]}{\pt^{\ell}},
\end{equation}
where $\sum\pt^\text{charged}$, $\sum\pt^\text{neutral-had}$, and $\sum\pt^{\gamma}$ are the scalar sums of the transverse momenta of charged hadrons originating from the PV, neutral hadrons and photons, respectively, and $\sum\pt^\mathrm{PU}$ is the sum of transverse momenta of charged hadrons not originating from the PV. In this analysis, the signal muon candidates are required to have $\mathrm{I}_{\text{rel}}^{\ell}<0.15$ and to pass the ``tight'' identification criteria described in Ref.~\cite{CMS:2018rym}. The combined identification and isolation efficiency varies between 90 and 99\% in the \pt range 20--200\GeV.

Hadronic decays of \PGt leptons are reconstructed using the ``hadrons-plus-strips'' algorithm~\cite{Chatrchyan:2012zz,Khachatryan:2015dfa,Sirunyan:2637646}, which classifies individual hadronic decay modes of the \PGt by combining charged hadrons from the PF reconstruction with neutral pions. The latter are reconstructed by clustering electrons and photons into rectangular strips, which are narrow in $\eta$ but wide in the $\phi$ dimension. The spread in $\phi$ accounts for photons originating from neutral pion decays, which convert into electron-positron pairs while traversing the silicon tracker. These particles are bent in opposite directions in $\phi$ by the magnetic field, and may further emit bremsstrahlung photons before reaching the ECAL. The decay modes considered in this analysis produce one charged pion or kaon plus zero, one or two neutral pions, or three charged pions or kaons plus zero or one  neutral pion. The  \textsc{DeepTau} algorithm identifies \tauh objects using a convolutional neural network combining information from high-level tau lepton observables together with low-level features obtained from the silicon tracker, the ECAL and HCAL, and the muon detectors for all the PF candidates, electrons, and muons that are reconstructed inside the tau lepton isolation cone ($\Delta R<0.5$). Three different discriminants are employed in order to distinguish \tauh candidates from jets (``\textsc{DeepTauVsJets}''), electrons (``\textsc{DeepTauVsEle}'') and muons (``\textsc{DeepTauVsMu}''). Different working points are used in each analysis channel, as summarized in Table~\ref{tab:selections}.

\begin{table*}[htb]
    \centering
    \topcaption{Summary of selections applied to the $\PGt\PGt$ candidate pair. Trigger thresholds in parentheses refer to the 2017--2018 data-taking period when different from 2016.}
    \setlength{\tabcolsep}{7pt}
    \renewcommand{\arraystretch}{1.2}
    \begin{tabular}{ll}
        \hline
        \multirow{5}{*}{Online \pt trigger thresholds}     & single-$\Pe$: $\pt>25\ (32)\GeV$ \\
                                                           & cross-$\Pe$: electron $\pt>24\GeV$, \tauh $\pt>30\GeV$ \\
                                                           & single-$\Pgm$: $\pt>22\ (24)\GeV$ \\
                                                           & cross-$\Pgm$: muon $\pt>19\ (20)\GeV$, \tauh $\pt>20\ (27)\GeV$ \\
                                                           & ditau: $\pt>35\GeV$, ditau VBF: $\pt>20\GeV$ \\
        \hline
        \multirow{2}{*}{Offline \pt thresholds}            & online threshold +1\GeV (electrons and muons) \\
                                                           & online threshold +5\GeV (\tauh candidates) \\
        \hline
        \multirow{2}{*}{$\eta$ thresholds}                 & electrons and muons: $\abs{\eta}<2.1$ \\
                                                           & tau: $\abs{\eta}<2.1$ (2.3) for ditau and cross (single) triggers \\
        \hline
        \multirow{2}{*}{Lepton ID and isolation}           & tight electron BDT ID + isolation \\
                                                           & tight muon ID and isolation \\
        \hline
        \multirow{3}{*}{\tauh ID (\eleth, \muth channels)} & medium \textsc{DeepTauVsJet} \\
                                                           & tight \textsc{DeepTauVsMu} \\
                                                           & very-loose \textsc{DeepTauVsEle} \\
        \hline
        \multirow{3}{*}{\tauh ID (\thth channel)}          & medium \textsc{DeepTauVsJet} \\
                                                           & very-loose \textsc{DeepTauVsMu} \\
                                                           & very-very-loose \textsc{DeepTauVsEle} \\
        \hline
        \multirow{2}{*}{Distance to PV}                    & $\abs{d_{xy}}<0.045\unit{cm}$ (electrons and muons only) \\
                                                           & $\abs{d_z}<0.2\unit{cm}$ \\
        \hline
        Pair selections                                    & opposite-sign, $\Delta R>0.5$ \\
        \hline
    \end{tabular}
    \label{tab:selections}
\end{table*}

Jets are reconstructed offline from PF candidates, clustered using the anti-\kt algorithm~\cite{Cacciari:2008gp,Cacciari:2011ma} with distance parameters of 0.4 and 0.8, respectively, denoted as ``AK4'' and ``AK8'' jets. The constituents of the AK8 jets are reclustered using the Cambridge--Aachen algorithm~\cite{Dokshitzer:1997in,Wobisch:1998wt}; the ``modified mass drop tagger'' algorithm~\cite{Dasgupta:2013ihk,Butterworth:2008iy}, also known as the ``soft-drop'' algorithm, is applied to remove soft, wide-angle radiation from the jet. The jet momentum is determined as the vectorial sum of all particle momenta in the jet, and is found from simulation to be, on average, within 5--10\% of the true momentum over the entire \pt spectrum and detector acceptance. Pileup collisions can give rise to jets not coming from the hard-scattering process (pileup jets) or contribute additional tracks and calorimetric energy depositions, increasing the apparent jet momentum. For AK4 jets, to mitigate this effect, tracks identified to be originating from pileup vertices are discarded and an offset is applied to correct for remaining contributions~\cite{Sirunyan:2020foa}. For AK8 jets, the pileup-per-particle identification algorithm~\cite{Bertolini:2014bba} is used to mitigate the effect of pileup at the reconstructed particle level, making use of local shape information, event pileup properties, and tracking information. Jet energy corrections are derived from simulation studies so that the average measured energy of jets becomes identical to that of particle-level jets. In situ measurements of the momentum balance in dijet, $\gamma+\mathrm{jets}$, $\PZ+\mathrm{jets}$, and multijet events are used to determine any residual differences between the jet energy scale in data and in simulation, and appropriate corrections are applied~\cite{Khachatryan:2016kdb,CMS-DP-2021-033}. Jets originating from \PQb{} quarks are identified using the \textsc{DeepJet} algorithm~\cite{Bols_2020}, a deep neural network combining secondary vertex properties, track-based variables, and PF jet constituents (neutral- and charged-particle candidates). Different thresholds applied to the output of the \textsc{DeepJet} algorithm result in working points characterized by their \PQb{} tagging accuracy; in this analysis the ``loose'' and ``medium'' working points are used, corresponding to an efficiency (misidentification rate for light-flavor and gluon jets) of 94 (10)\% and 84 (1)\%, respectively~\cite{Bols_2020}.

As mentioned in the introduction, this analysis considers three decay modes of the $\PGt\PGt$ system: \muth, \eleth, and \thth.
Different trigger strategies are employed for each of these channels. In the semileptonic ones, single-lepton and cross-lepton triggers are used. The former requires the presence of an isolated electron or muon only, while the latter also requires an additional hadronically decaying tau lepton. The use of cross-triggers represents the main change in the trigger strategy with respect to the result previously published on 2016 data~\cite{Sirunyan:2273798} and improves the analysis sensitivity, allowing the \pt threshold on electrons and muons to be lowered, as described in Table~\ref{tab:selections}. In the \thth channel, triggers requiring the presence of a pair of hadronically decaying \PGt leptons with $\pt>35\GeV$ are employed. A VBF ditau trigger was introduced in 2017 to extend the analysis phase space, targetting events containing two hadronically decaying \PGt leptons with $\pt>20\GeV$ and two jets, one with $\pt>115\GeV$ and the other with $\pt>45\GeV$. The online trigger thresholds for each electron, muon or tau lepton are reported in Table~\ref{tab:selections}. Compared to the online selection, the offline thresholds are increased by 1\GeV for electrons and muons and by 5\GeV for \tauh candidates: this is done to assure that the trigger efficiency reaches its plateau value.

Events are classified in three different $\PGt\PGt$ decay channels based on the information from the \PGt lepton candidates reconstructed offline. Selected signal events are required to have at least one \tauh candidate. An event is classified as \muth if a muon passing the selections detailed in Table~\ref{tab:selections} is found, otherwise it is classified as \eleth if an electron passing the selections is found: if neither a muon nor an electron are found, the event is classified as \thth if a second \tauh is present. In the \thth channel, if not stated otherwise, the former \tauh refers to the more isolated one.

Each reconstructed electron, muon, or tau lepton candidate is required to have a distance between the track of the candidate and the PV in the transverse and longitudinal planes satisfying $\abs{d_{xy}}<0.045\unit{cm}$ (electrons and muons only) and $\abs{d_z}<0.2\unit{cm}$. The two objects forming a pair must have opposite charge and be separated by $\Delta R>0.5$. To reduce $\PZ/\PGg^*\to\Pell\Pell$ contamination, a third-lepton veto is applied, by discarding events where additional electrons or muons are present.

Finally, events selected with the above described criteria are required to have at least two additional AK4 jets with $\pt>20\GeV$, $\abs{\eta}<2.4$, and $\Delta R(j,\Pell)>0.5$, where \Pell represents each of the two leptons forming the $\PGt\PGt$ candidate pair. Studies on the previous published \bbtt result~\cite{Sirunyan:2273798} highlighted the selection of the $\PH\to\PQb\PQb$ candidate as being one of the least efficient. A new \PQb{} jet assignment algorithm (referred to as \HH-bTag) has therefore been developed, based on a neural network architecture. The input features of the network include the \PQb{} jets and $\PH\to\PGt\PGt$ system kinematic variables, the \textsc{DeepJet} score of the \PQb{} jet candidates, and their angular separation with respect to the $\PH\to\PGt\PGt$ candidate. For each event, all possible \PQb{} jet candidates are assigned a score by the \HH-btag algorithm; the two jets with the highest scores are taken as those originating from the decay of the \PH{}. Two additional jets are looked for amongst all the remaining jets with $\pt>30\GeV$ and $\abs{\eta}<4.7$. Among all the additional jets in the event, all possible pairs are built and that with the highest invariant mass ($m_{\mathrm{jj}}$) is chosen as the VBF candidate pair. In case the event was selected by the VBF ditau trigger, and not by the other ditau triggers, due to the lower \pt thresholds on \tauh candidates, three additional requirements are then made: $m_{\mathrm{jj}}>800\GeV$, $\pt>140\GeV$ and 60\GeV for the leading and subleading VBF jet candidates.

The selections on leptons and jets described in this section and in Table~\ref{tab:selections} lead to a final efficiency and acceptance product of 5.5 (3.4)\% for the \ggF{} (VBF) SM signal events.

\section{Event categorization and discriminating variables}
\label{sec:categories}
Events selected as described in Section~\ref{sec:objects} are classified in eight mutually exclusive categories, in order to maximize the sensitivity to both the ggF and VBF HH searches.

In order to categorize signal events produced through the VBF mechanism, a selection (denoted ``VBF inclusive'') is introduced by requiring that two VBF jet candidates, as described in Section~\ref{sec:objects}, can be identified in the event and that they further pass two requirements (jointly referred to as the ``VBF tag'') designed to reduce the background contamination: $m_{\mathrm{jj}}>500\GeV$ and a separation in pseudorapidity ($\Delta\eta$) between the two VBF jet candidates larger than 3. Because of the \ggF{} signal contamination in the VBF inclusive selection, in order to enhance the analysis sensitivity, a multiclassification approach is introduced that allows to create two additional signal categories (\textit{classVBF} for the \HH VBF events and \textit{classGGF} for the \ggF{} contribution), as well as three background-enriched ones (\textit{classttH} for \ttH{} events, \textit{classTT} for the \ttbar contamination, and \textit{classDY} for the $\PZ/\PGg^*\to\Pell\Pell$ contribution), used to constrain the systematic uncertainties affecting these processes. Each event is assigned to the category for which it has the highest multiclassifier score. The score represents the probability of originating from the specific process denoted by the category name.

Events without two VBF jet candidates identified, or with two VBF jet candidates identified but not passing the VBF tag selection, are classified exploiting the differences in Lorentz boost regimes of \PQb{}\PAQb{} production in order to increase the analysis sensitivity. If the two \PQb{} jets have $\Delta R(\PQb,\PQb)>0.8$, jets are reconstructed as separated AK4 jets; on the other hand, if they have $\Delta R(\PQb,\PQb)<0.8$ then they are merged and reconstructed as a single AK8 jet. In the first case, the jets are referred to as ``resolved'', in the latter case as ``boosted''. To be categorized as boosted, events are required to have a minimal AK8 jet mass of 30\GeV, a distance between the two subjets and the previously selected AK4 jets of $\Delta R<0.4$, and both the subjets must pass the loose \PQb{} tagging working point. All events not satisfying these requirements are categorized as resolved. In addition, resolved events are split into two categories depending on the \PQb{} tag multiplicity: events where only one jet passes the medium \PQb{} tagging working point (\textit{res1b}), and events where both jets pass this working point (\textit{res2b}).

After object selection and event categorization, the kinematic information in the events is used to introduce additional requirements on the invariant masses of the \PQb{} jet candidates ($m_{\PQb\PQb}$) and the tau lepton candidates ($m_{\PGt\PGt}$): the latter is reconstructed using the \textsc{SVfit} algorithm~\cite{1742-6596-513-2-022035}, a dynamic likelihood technique combining the kinematical properties of the two visible lepton candidates and the \ptmiss in the event.
Events satisfying the invariant mass requirements constitute the signal region (SR) of the analysis.
For the resolved categories the selection is:
\begin{equation}
\frac{\left(m_{\PGt\PGt} - 129\GeV \right)^2}{\left( 53\GeV \right)^2} + \dfrac{\left(m_{\PQb\PQb} - 169\GeV \right)^2}{\left( 145\GeV \right)^2} < 1,
\label{eq:resolvedMcut}
\end{equation}
while for the \textit{boosted} category, because of its different kinematical properties, the selection is:
\begin{equation}
\frac{\left(m_{\PGt\PGt} - 128\GeV \right)^2}{\left( 60\GeV \right)^2} + \dfrac{\left(m_{\PQb\PQb} - 159\GeV \right)^2}{\left( 94\GeV \right)^2} < 1.
\label{eq:boostedMcut}
\end{equation}
No invariant mass requirement is applied to the VBF multicategories.

In Eqs. (\ref{eq:resolvedMcut}) and (\ref{eq:boostedMcut}), the mass offsets (numerator) and resolutions (denominator) are chosen using a random search that minimizes
the background acceptance for a signal efficiency larger than 90\%. Contrary to what was done in Ref.~\cite{Sirunyan:2273798}, the purpose of the invariant mass selection
is no longer to maximize the signal significance, but rather to remove significantly outlying background events in regions where no signal is expected,
while the discrimination of \HH events from the background is left to a specifically designed neural network, as described in the next paragraph.

The eight orthogonal categories (\textit{res2b}, \textit{res1b}, \textit{boosted}, \textit{classVBF}, \textit{classGGF}, \textit{classttH}, \textit{classTT} and \textit{classDY}) obtained are used for the signal extraction via a deep neural network developed to identify $\HH\to\bbtt$ events. In the following, this discriminating algorithm will be referred to simply as DNN. 
A single training is performed with events from all years, channels and categories, obtained from the simulated background and the SM \HH{} signal samples as described in Section~\ref{sec:datasets}. The scope of the training is to classify events as originating either from signal or background processes by assigning them a single prediction. Predictions closer to zero indicate ``background-like'' events, while predictions closer to one indicate ``signal-like'' events.  The final DNN distributions used in the signal extraction fit are obtained by inferring predictions of the trained network separately in each of the eight orthogonal categories, split by channel and year.

The DNN is composed of two discriminators, each consisting of ten neural networks trained via ten-fold stratified cross-validation of the training set. The architecture of the networks is the same as the optimal architecture developed in Ref.~\cite{higgsml_lumin}, with the addition of increased overfitting supervision~\cite{SDR,dropout}. The approach used to train the DNN~\cite{perceptron,Linnainmaa_70,Werbos_81,backprop} follows that used in the CMS \HH{} High-Luminosity LHC projection analysis for the \bbtt final state~\cite{CMS-HL-LHC,Higgs-HL-LHC}. In this approach, the simulated events are split evenly into two subsets and each discriminator of the DNN is trained on a different subset. At the inference stage, the discriminators are used in the subset of events on which they were not trained. Thus, all of the simulated events are used for inference, reducing the associated statistical uncertainty on the sample density distributions, while not being subject to a systematic bias in the predictions that would otherwise arise if the discriminators were applied to events on which they were actually trained. It should be noted that this approach does not provide an explicit validation sample, and does therefore not permit any fine-tuning of hyper-parameters without the risk of biasing the model. Nevertheless, the lack of tuning is not expected to significantly impair the performance of the DNN~\cite{higgsml_lumin}. A total of 26 features (selected from a starting pool of over 100) are used as an input to the DNN. Besides the categorical features, such as the year of data taking, the $\PGt\PGt$ decay mode, and the number of VBF jet candidates, among the most important features are the \textsc{DeepJet} scores of the \PQb{} jets, binned in working points, the invariant masses of the \HH{}, $\PGt\PGt$, and $\PQb\PQb$ systems, and several kinematic variables, such as visible momenta and distances in the $\eta-\phi-$plane between reconstructed particles, describing the spatial distribution of the physics objects in the event.

\section{Background modeling}
\label{sec:background} 
The main background sources contaminating the SR are \ttbar production, $\PZ/\PGg^*\to\Pell\Pell$ production, and QCD multijet events; all three are modeled using specific approaches. The productions of single top, single \PH{}, \PW{} boson in association with jets, diboson, triboson, and \ttbar pairs in association with a single boson or a pair of vectors bosons are instead estimated from simulation, as described in Section~\ref{sec:datasets}.
For all backgrounds, except multijet, both events with real hadronically decaying tau leptons and with lepton or quark or gluon jets misidentified as \tauh candidates
are taken from simulation. The fraction of misidentified \tauh candidates in the dominant background (\ttbar) is lower than 20\% in the analysis SR.
No specific correction is applied to compensate for potential differences between simulation and data for fake \tauh candidates, but a systematic uncertainty
is added to the fit model as described in Section~\ref{sec:systematics}.

The \ttbar background contribution is modeled relying on simulation. The normalization of the background shows a disagreement with respect to the observed data,
thus a correction scale factor (\textit{ttSF}) is fitted from a \ttbar-enriched control region (CR).
Three CRs are defined, one for each year, to contain \eleth, \muth, and \thth events that satisfy the \textit{res2b} selections
outlined in Section~\ref{sec:categories} and pass a mass requirement obtained by inverting, among the resolved categories, the one defined in Eq. (\ref{eq:resolvedMcut}).
The contamination from other backgrounds in the \ttbar-enriched CR is less than 9\% for all three years.
The \textit{ttSF}s resulting from the fit are 0.908, 0.988, and 0.966 for 2016, 2017, and 2018, respectively, with an uncertainty, also coming from the fit,
that ranges between 0.6 and 0.9\%.

The $\PZ/\PGg^*\to\Pell\Pell$ background contribution is modeled relying on simulation. Scale factors for the simulation normalization are taken from 18 CRs containing events with two isolated muons compatible with the $\PZ\to\PGm\PGm$ decay. The same jet selection criteria as in the SR are applied in the CRs. The CRs are defined as follows: the data are first split into three CRs based on the number of \PQb-tagged jets (0, 1, or 2); each CR so obtained is further split into six CRs based on the \pt of the reconstructed \PZ{} candidate. Simulated events are split, obtaining 18 templates based on the number of \PQb{} partons at the generator level and the \pt of the generator-level \PZ{} boson; one additional template, combining all other backgrounds, is also considered. The scale factors are obtained performing a simultaneous fit of the di-muon mass in the CRs using the 19 templates, where their normalizations are kept floating.

The multijet background contribution is determined from data in jet-enriched regions where the tau lepton pair requirements are inverted. Two different shapes are obtained in regions where either the pair charge or the \tauh isolation requirement is inverted (in the \thth final state, only the isolation requirement on the lowest \pt \tauh candidate is inverted) and all the other selections are applied as in the SR. The final multijet background template is then defined as the mean of these two shapes. The yield is obtained as the product of the yields in the two regions defined above divided by the yield obtained in the region where both the charge and isolation requirements are inverted. The contributions of other backgrounds, based on predictions as outlined above, are subtracted in all these regions.

\begin{figure*}[!htb]
\centering
\includegraphics[width=0.48\textwidth]{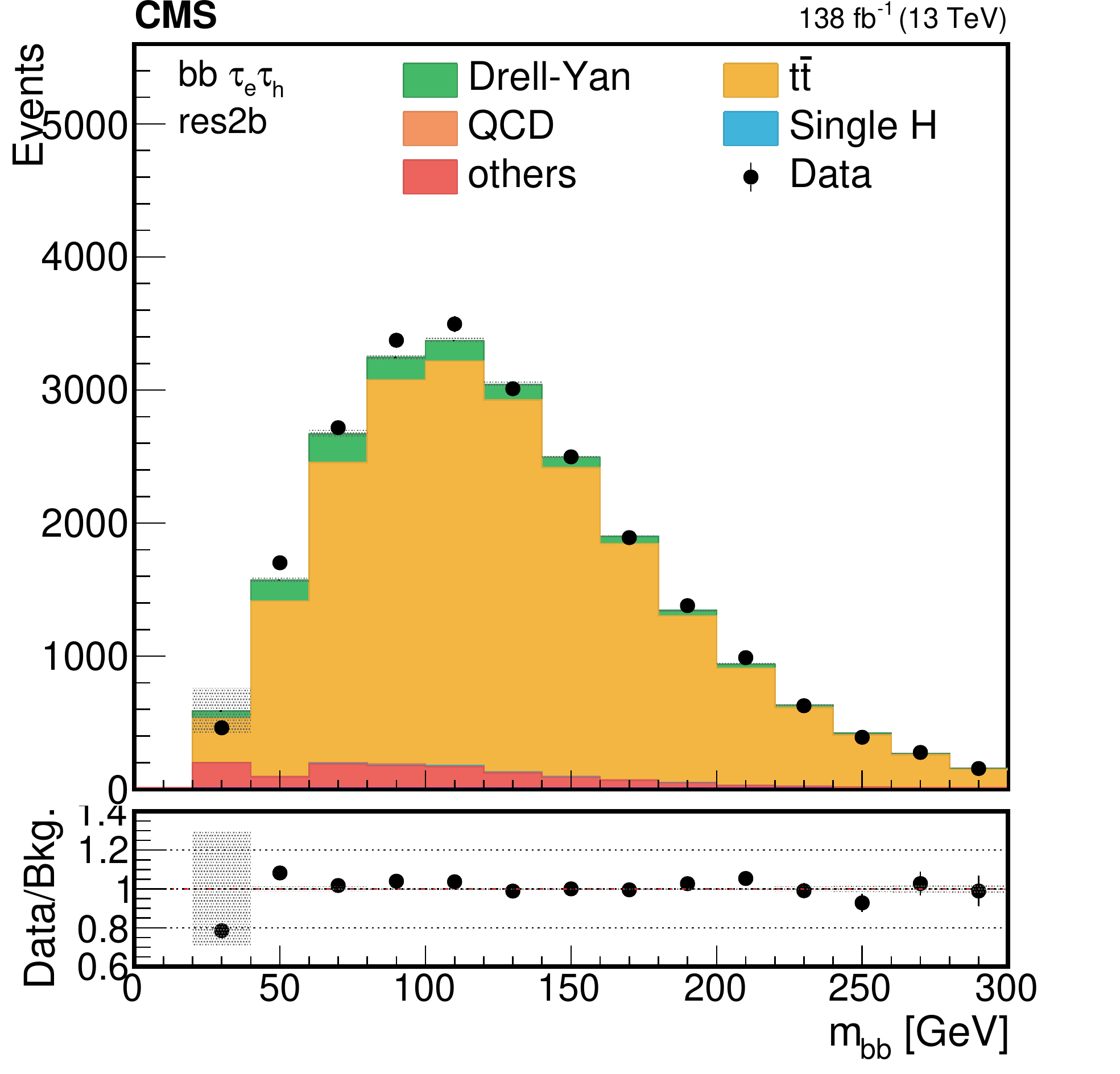}
\includegraphics[width=0.48\textwidth]{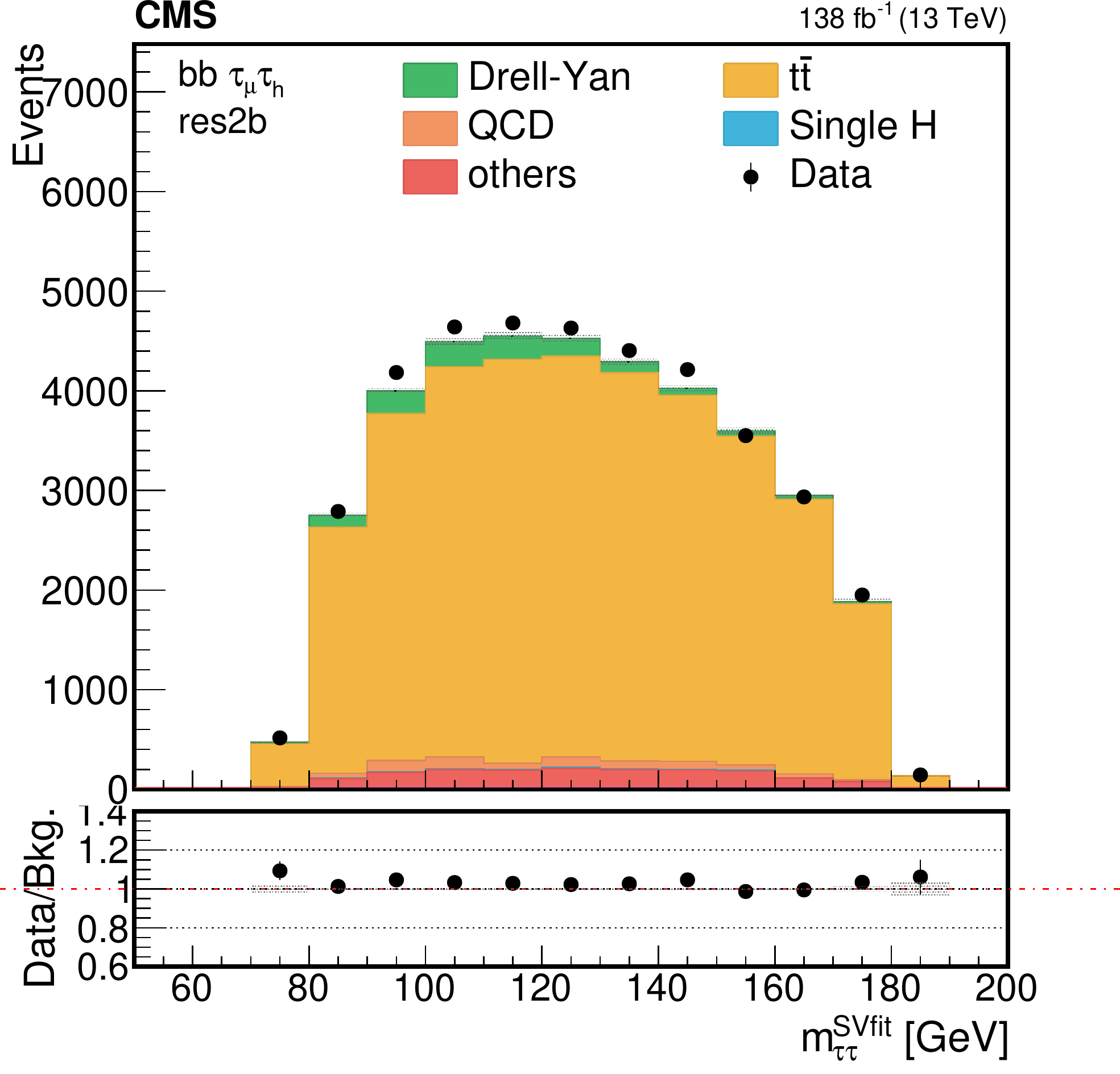}\\
\includegraphics[width=0.48\textwidth]{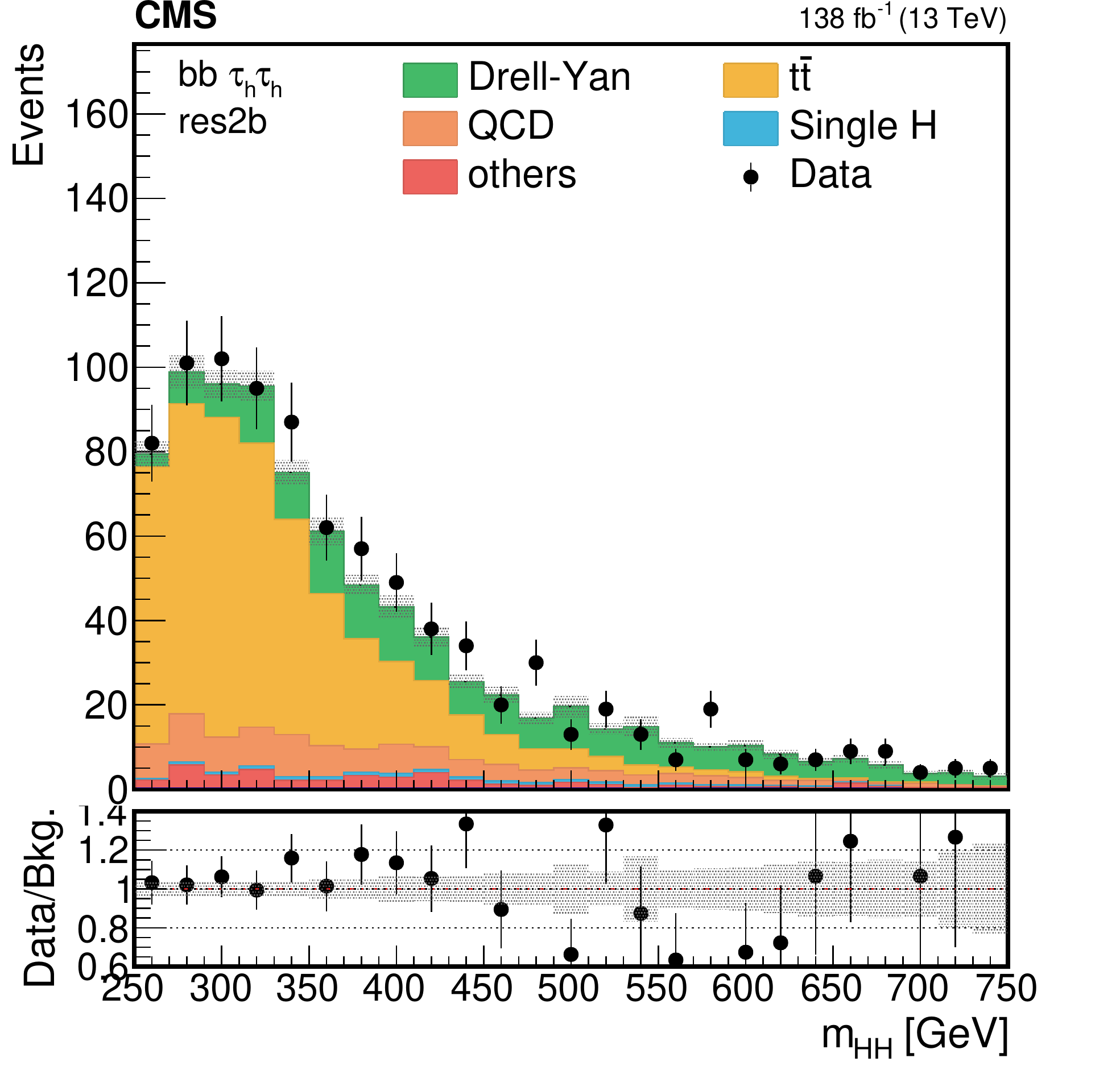}
\caption{
    Distributions of the reconstructed mass of the $\PQb\PQb$ (upper left), $\PGt\PGt$ (upper right) and \HH (lower) pairs in the most sensitive
    category of the analysis (\textit{res2b}). Events are shown in the \eleth (upper left), \muth (upper right) and \thth (lower) channels for
    the full 2016--2018 data set, after the selection on the reconstructed masses of the $\PGt\PGt$ and $\PQb\PQb$ pairs, as described in Section~\ref{sec:categories}.
    The shaded band in the plots represents the statistical uncertainty only.
}
\label{fig:mass_distributions}
\end{figure*}

A comparison of data and simulation prediction, corrected as described in this Section, for three of the most important input features of the
DNN described in Section~\ref{sec:categories} is shown in Fig.~\ref{fig:mass_distributions}.

\section{Systematic uncertainties}
\label{sec:systematics} 
In this analysis, we include various sources of systematic uncertainty that originate from a limited knowledge of the background and signal processes, discrepancies between simulation and data, and imperfect knowledge of the detector response. They are categorized as ``normalization'' and ``shape'' uncertainties: while the first affect only the total event yield of the processes, the latter affect also the distribution of the events. The uncertainties described in this Section are introduced as nuisance parameters in the maximum likelihood fit described in Section~\ref{sec:results}, with log-normal (Gaussian) priors for normalization (shape) uncertainties.

The following normalization uncertainties are considered.
\begin{itemize}
\item Different sources of uncertainty in the integrated luminosity measurement are considered and treated as correlated. Their values are obtained from dedicated Van-der-Meer scans and the stability of detector response during data taking. The integrated luminosities for the 2016, 2017, and 2018 data-taking years have 1.2--2.5\% individual uncertainties~\cite{CMS-LUM-17-003,CMS-PAS-LUM-17-004,CMS-PAS-LUM-18-002}, while the overall uncertainty for the 2016--2018 period is 1.6\%. These uncertainties are applied only to the signals and to background processes estimated from simulation. Since the normalizations of the \ttbar, $\PZ/\PGg^*\to\Pell\Pell$, and multijet backgrounds are obtained from data, they are not subject to the integrated luminosity uncertainties.
\item Electron and muon reconstruction, isolation, and identification uncertainties are determined from the simulation-to-data scale factors; a value of 1\% for both electrons and muons is obtained~\cite{CMS:2020uim,CMS:2018rym}. An additional uncertainty of 3 (15)\% for tau leptons with $\pt<100\GeV$ ($>100\GeV$) is added in the $\thth$ channel~\cite{CMS:2022prd}.
\item During the 2016--2017 data taking, a gradual shift in the timing of the inputs of the ECAL L1 trigger in the region at $\abs{\eta}>2.0$ caused a trigger inefficiency. For events containing an electron (a jet) with $\pt>50\text{ (100)}\GeV$, in the region $2.5<\abs{\eta}<3.0$ the efficiency loss is approximately 10--20\%, depending on \pt, $\eta$, and time. Correction factors were computed from data and applied to the acceptance evaluated by simulation. An uncertainty of 2\% is assigned to this effect.
\item The uncertainty in the pileup reweighting technique is estimated by varying the values of the applied pileup weights by their uncertainty. The resulting systematic uncertainty is estimated to have a value of 1\% and it is correlated among all channels and categories in each year.
\item The normalization of the \ttbar background is taken from a fit to a CR per year, as described in Section~\ref{sec:background}. The uncertainty in the scale factors obtained in these CRs is purely statistical and is always below 1\%.
\item The normalization of the $\PZ/\PGg^*\to\Pell\Pell$ background is taken from a fit to 18 CRs per year, as described in Section~\ref{sec:background}. The uncertainties in the scale factors obtained in these CRs are propagated to the SR taking into account their correlation and range from 0.1 to 60\% depending on the year and CR considered.
\item The multijet background contribution is determined from data in jet-enriched regions, as described in Section~\ref{sec:background}. Two normalization uncertainties are derived in order to take into account the event yield statistical fluctuations in these CRs and their dependence on the \tauh isolation requirement used to define them.
\item The uncertainties in the normalizations of the backgrounds modeled relying solely on the simulated events range from 2 to 10\%.
\item The theoretical uncertainty in the cross section of \HH production: via \ggF{}: $^{+6\%}_{-23\%}$ (scale $+\ m_{\PQt}\text{)}$, $\pm3\%\text{ (PDF}+\alpS\text{)}\unit{fb}$~\cite{Grazzini_2018}; and via VBF: $^{+0.03}_{-0.04}\%$ (scale) and $\pm2.1\%$ (PDF $+\,\alpS$)~\cite{PhysRevD.98.114016}. These uncertainties are only considered when upper limits are quoted with respect to the SM and for the likelihood scans, they are not included for the upper limits on the cross section.
\item The theoretical uncertainty in the \PH{} branching fractions~\cite{YR4} amount: for the decay to $\PQb\PAQb$, to $\pm0.65\%$ (theory), $^{+0.72}_{-0.74}\%$ ($m_{\PQq}$), $^{+0.78}_{-0.80}\%$ (\alpS), where $m_{\PQq}$ is the quark mass; and for the decay to tau leptons to $^{+1.16}_{-1.17}\%$ (theory), $^{+0.98}_{-0.99}\%$ ($m_{\PQq}$), $0.62\%$ (\alpS).
\item The uncertainty in the modeling of the VBF signal in {\PYTHIA}8 is estimated through samples generated setting the dipole recoil option, which affects the initial-state parton showers, to ON~\cite{dipoleShoweStudies}. The ratio of integrated yields in the dipole recoil ON/OFF samples is taken as the uncertainty: it varies from 10\% for largely populated categories such as \textit{classVBF}, to 70\% for categories with poor VBF signal contribution, such as \textit{classttH}.
\end{itemize}

The following shape uncertainties are considered.
\begin{itemize}
\item The uncertainty in the measurement of the energy of \tauh leptons~\cite{CMS:2022prd}. The uncertainties are derived by combining low- and high-\pt measurements in $\PZ\to\PGt\PGt$ and in the off-shell $\PW^{*}\to\PGt\PGn$ events. Four different uncertainties are included to take into account the different \tauh decay modes considered in this analysis. When considering the uncertainty for a particular decay mode, the shift is applied only to the \tauh candidates that are reconstructed with that particular decay mode, while all other \tauh candidates are left unchanged.
\item Uncertainties related to the calibration of jet energy scale (JES) and resolution (JER). For JES, 11 separate sources of uncertainty are included per year: those appearing in multiple years are treated as fully correlated, while those appearing only in one year are treated as uncorrelated. For JER, alternative templates are produced by shifting all the jet-related features. These shifts stem from the use of scale factors to smear the energy of simulated jets to match the observed energy resolution in data.
\item Separate uncertainties in the energy scale of electrons misidentified as \tauh candidates are provided to take into account two different decay modes, h$^{\pm}$ and h$^{\pm}\PGpz$. The uncertainty in the energy scale of muons misidentified as \tauh candidates is 1\%, uncorrelated across decay modes.
\item The uncertainties arising from the application of the \textsc{DeepTau} identification scale factors are determined using a tag-and-probe procedure as a function of the \tauh candidate \pt. Five uncertainties are computed for the identification of taus against jets and are calculated in \pt bins of the \tauh candidate. All these uncertainties are used in the \muth and \eleth channels, while, since in the \thth channel both leptons are required to have \pt above a threshold, only the highest \pt bin uncertainty can be applied. Two uncertainties are present in the identification of tau leptons against electrons, one for the barrel and one for the endcap, and are treated as uncorrelated across $\eta$ bins of the \tauh candidate.
\item The shape uncertainty in the multijet contribution is determined from the two alternative templates described in Section~\ref{sec:background}.
\item The uncertainties arising from the misidentification of jets as \tauh candidates are determined from \muth CRs defined inverting the charge sign requirement on the tau leptons and imposing that neither of the \PQb{} jet candidates pass the ``medium'' working point of the \textsc{DeepJet} algorithm. Six uncertainties, one for the barrel and one for the endcap for each year, are derived and treated as uncorrelated across years and detector regions.
\item Trigger efficiencies and scale factors are measured in a $\PZ\to\PGt\PGt\to\PGm\PGnGm\PGnGt\tauh\PGnGt$ enriched region with a tag-and-probe procedure and fitted separately for data and simulation. The uncertainty, determined as a function of the reconstructed \tauh candidate \pt, is taken from the fit. The scale factors are obtained from the data to simulation ratio and their uncertainties are propagated accordingly. Four uncertainties are included to take into account the different \tauh decay modes considered in this analysis. They are applied to the \tauh leg of each channel. Two additional trigger uncertainties are used to cover the cases where the tau lepton decays to an electron or a muon. Finally, one uncertainty is added for the \thth final state in 2017--2018 to take into account the jet legs scale factors of the VBF trigger.
\item The uncertainty in the \PQb{}-tagging efficiency. This takes into account the contamination from $\PQu\PQd\PQs\PQc\Pg$ ($\PQc\PQb$) jets in heavy- (light-) flavor regions, as well as statistical fluctuations in both data and simulation samples used for the computation of the  \PQb{}-tagging efficiency.
\item The uncertainties in the pileup jet identification scale factor as functions of \pt and $\eta$.
\end{itemize}

The largest sources of systematic uncertainties come from the imperfect knowledge of the \ggF{} \HH production cross section, the statistical fluctuations affecting the multijet background estimation, and the mismodeling of jet and \PGt leptons identification and reconstruction in simulated samples. The total impact of the systematic uncertainties on the result is approximately 15\%.

\section{Results}
\label{sec:results}
A binned maximum likelihood fit of the DNN prediction to the data is performed in eight categories per channel simultaneously
for all three years, for a total of 72 input distributions. The systematic uncertainties described in Section~\ref{sec:systematics}
are introduced as nuisance parameters in the likelihood function.
Upper limits at 95\% confidence level (\CL) on the \HH production cross section are set using the asymptotic modified frequentist method (asymptotic \CLs)~\cite{CLS1,CLS2,Cowan:2010js}.
We report here the expected and observed sensitivities for the inclusive (\ggF + VBF) \HH production cross section
as functions of $\kappa_{\lambda}$, whereas the expected and observed sensitivities for the VBF \HH production cross section are reported as functions of $\kappa_{2\PV}$. Coupling modifiers that are not part of the scanning procedure are fixed to unity, which corresponds to the SM prediction.

The binning schemes for the DNN prediction distributions used in the likelihood fit are chosen to minimise the expected upper limit, while keeping the number of bins the smallest possible and ensuring the stability of the fit. The binning schemes start from finely-binned histograms: bins are then merged using Bayesian optimization techniques to find the best performing scheme~\cite{Mockus1991}.  To reduce computational complexity, limits are minimized consecutively by combining one category after the other and freezing the binning for all previous histograms.  In order to avoid poor modeling of background yield or background uncertainty, each bin is required to contain at least one \ttbar and one $\PZ/\PGg^*\to\Pell\Pell$ event and to have a weighted background yield larger than 0.03. 

\begin{figure}[!htb]
\centering
\includegraphics[width=0.48\textwidth]{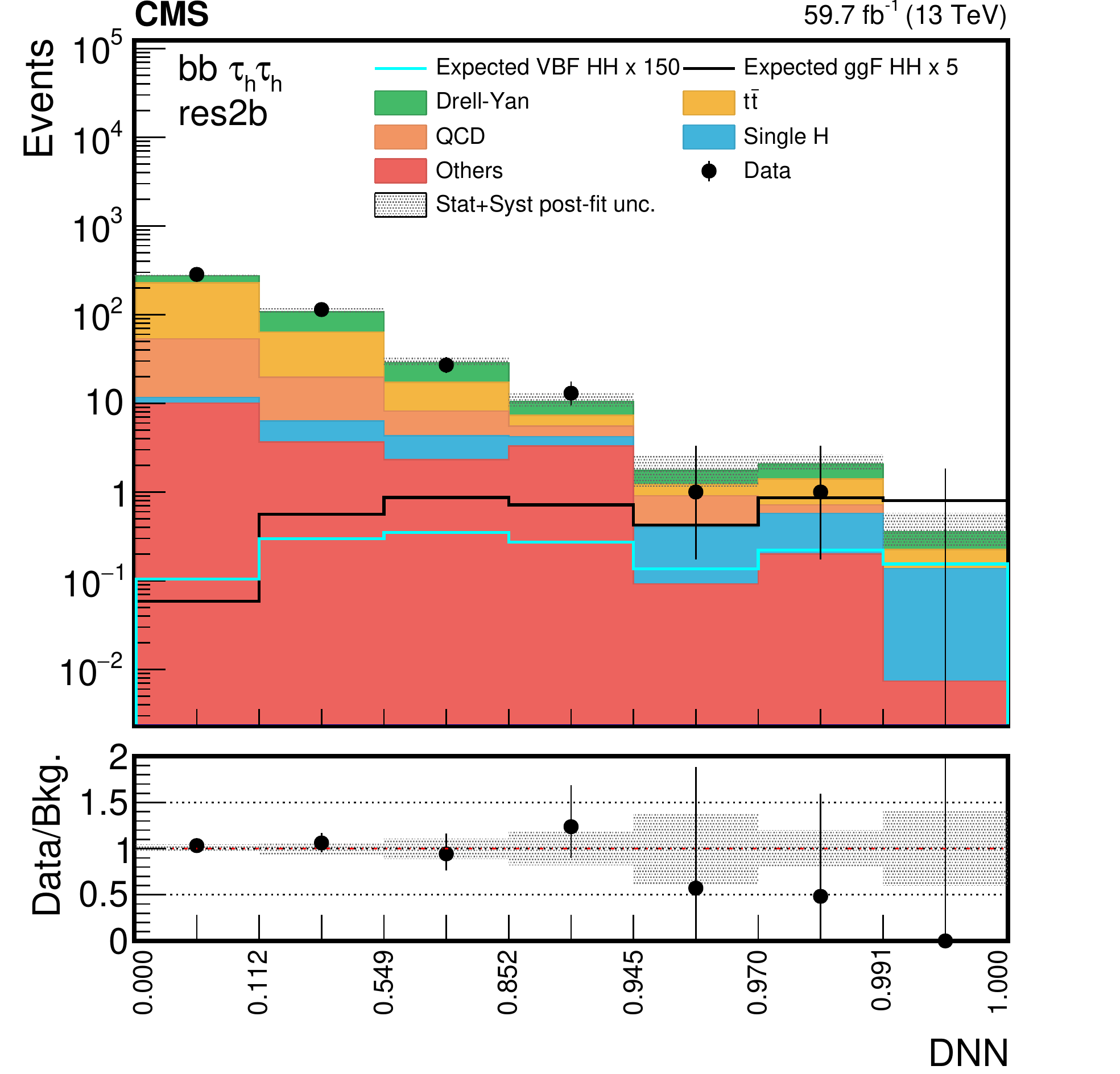}
\includegraphics[width=0.48\textwidth]{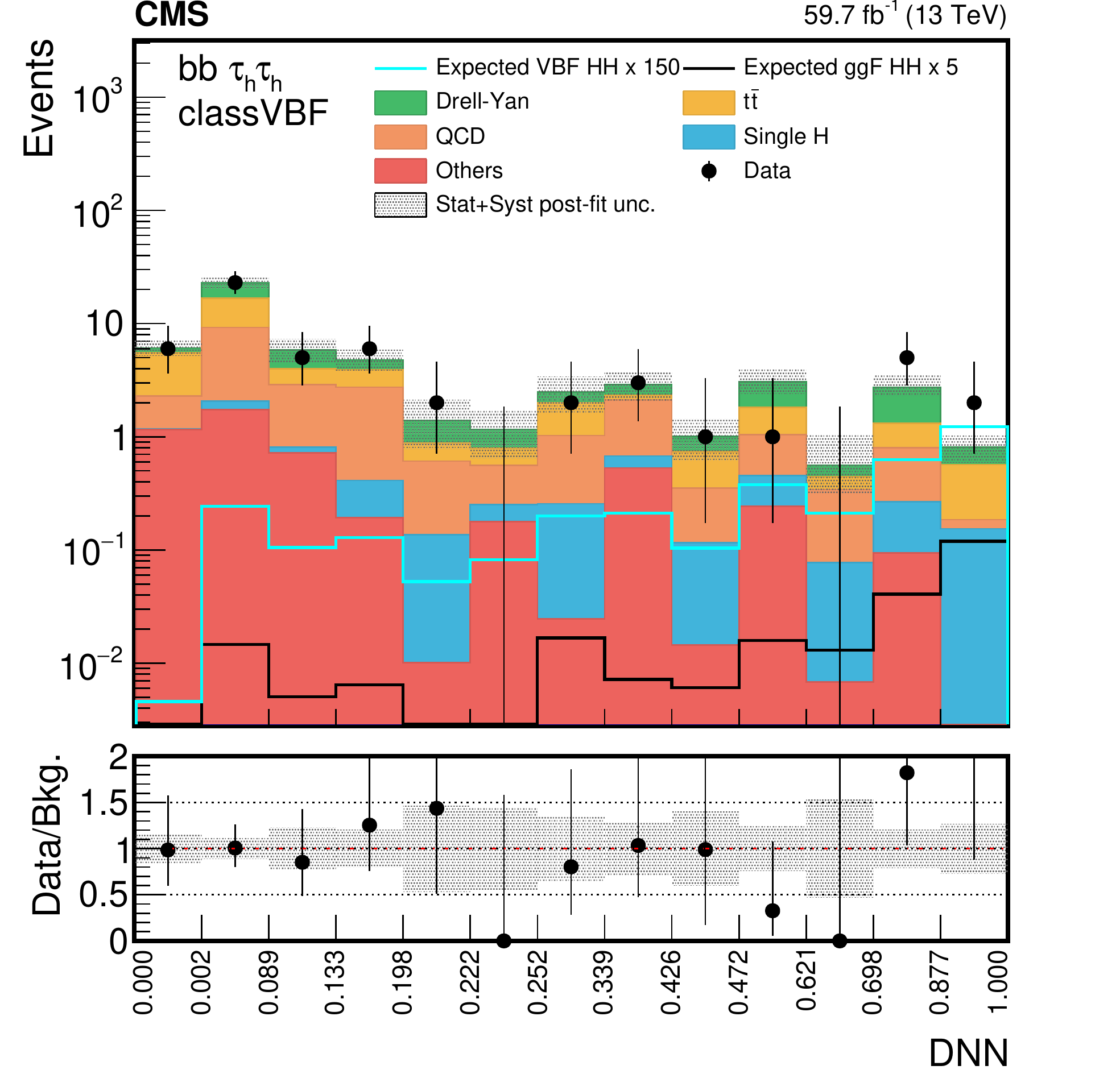}
\caption{The postfit DNN distributions in the \thth channel in 2018 for the most sensitive category in the \ggF{} (\cmsLeft) and VBF (\cmsRight) searches. The shaded band in the plots represents the statistical plus systematic uncertainty. The expected SM distributions for the \ggF and VBF HH signals are shown superimposed on the figures.}
\label{fig:scoresTauTau2018}
\end{figure}

The obtained postfit DNN distributions used for the maximum likelihood fit are exemplified in Fig.~\ref{fig:scoresTauTau2018},
where the distributions for the two most sensitive categories in the \ggF{} (\textit{res2b}) and VBF (\textit{classVBF}) searches in the \thth channel in 2018 are shown;
the distributions are plotted as a function of the DNN bin number for an easier visualization.

\begin{figure*}[!htb]
\centering
\includegraphics[width=0.48\textwidth]{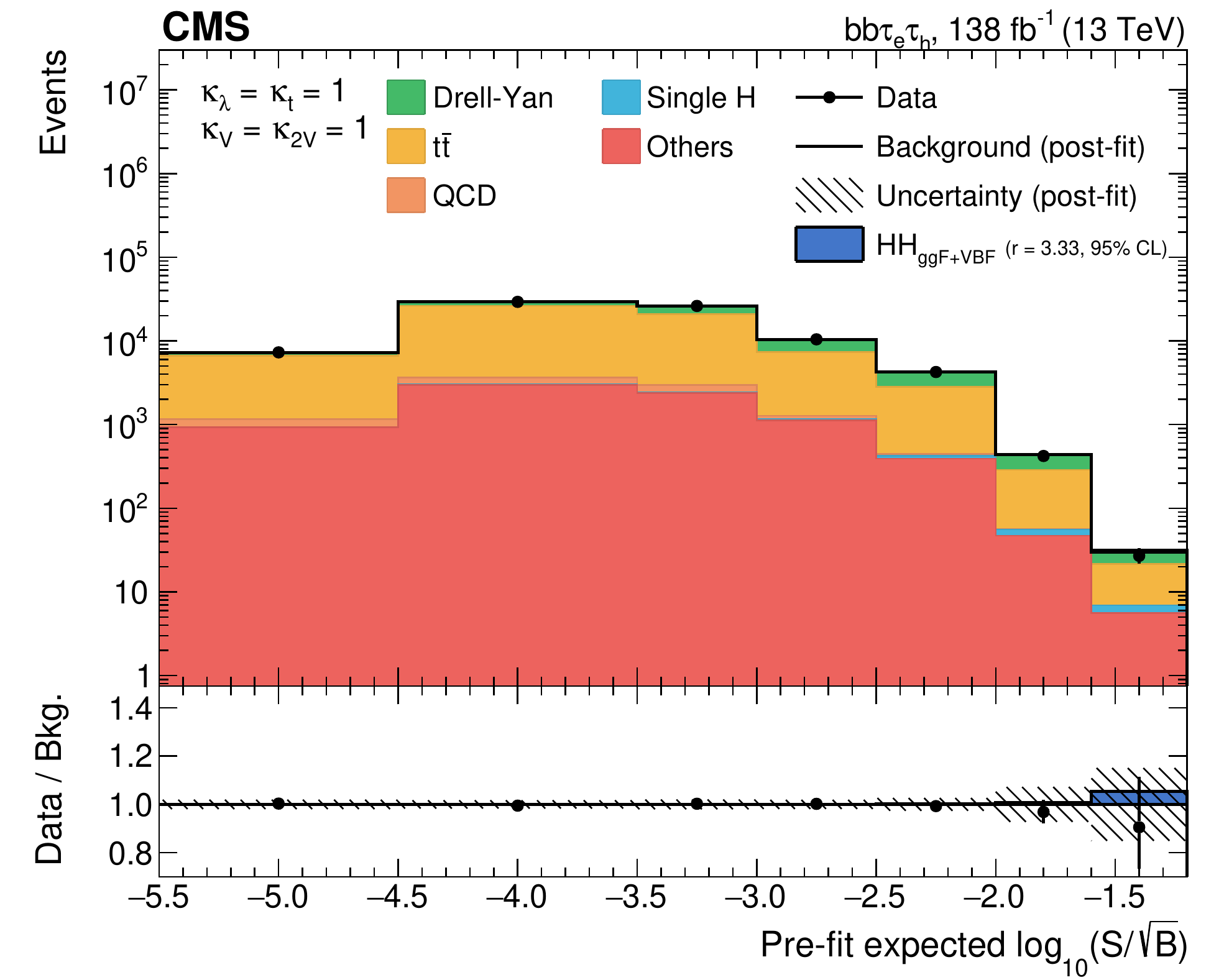}
\includegraphics[width=0.48\textwidth]{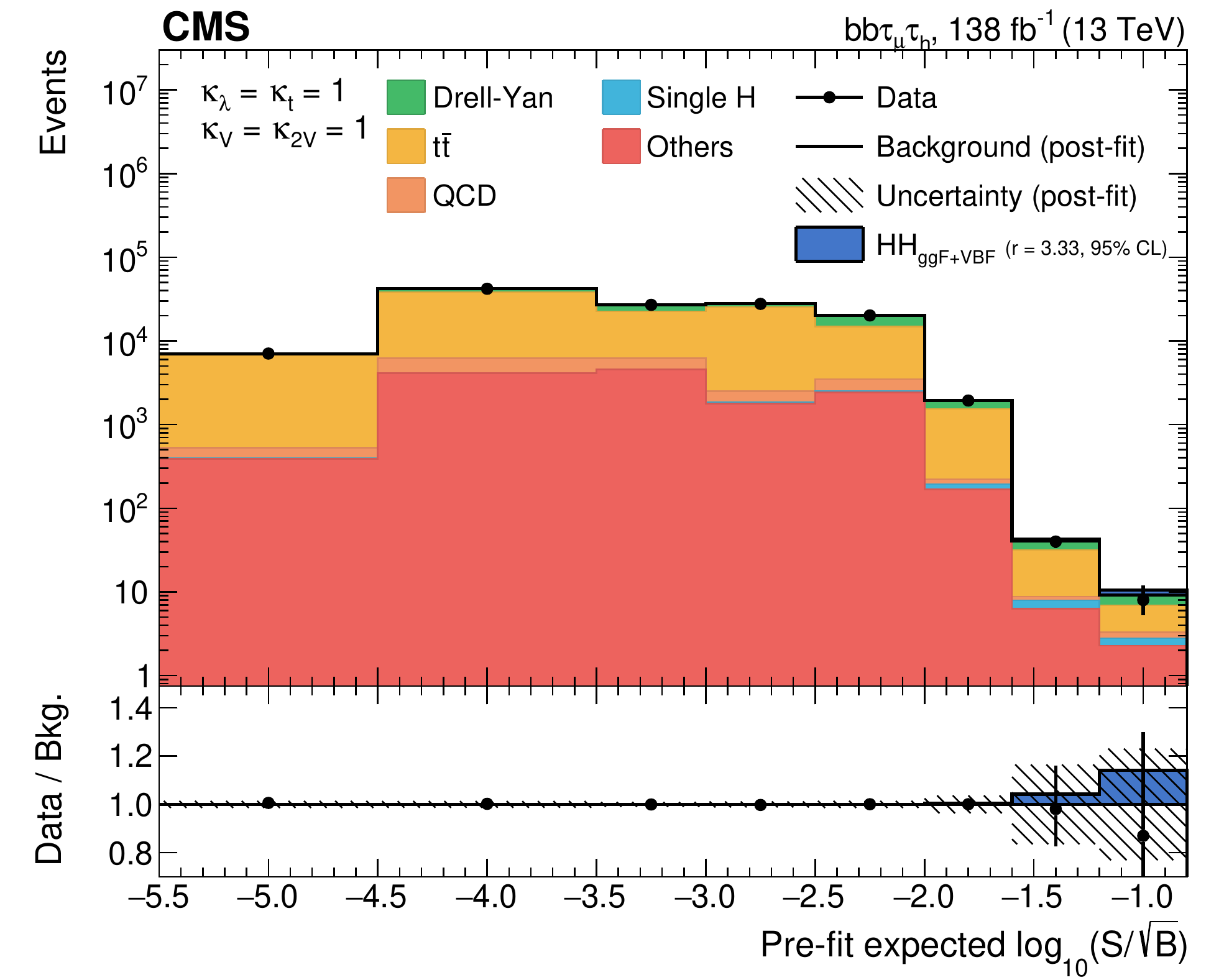}\\
\includegraphics[width=0.48\textwidth]{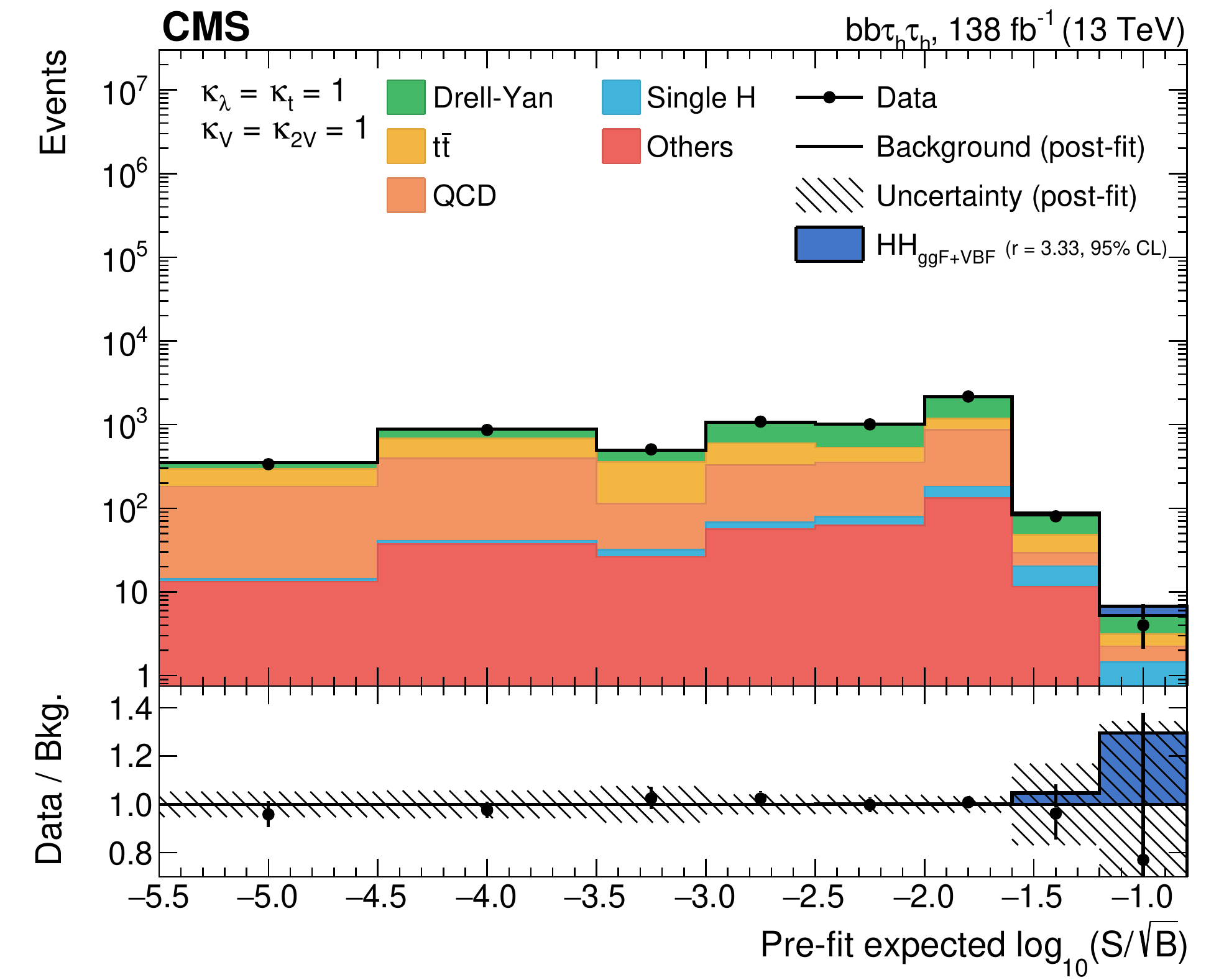}
\caption{
    Combination of bins of all postfit distributions, ordered according to the expected signal-to-square-root-background ratio,
    where the signal is the SM \HH signal expected in that bin and the background is the prefit background estimate in the same bin,
    separately for the \eleth channel (upper left), the \muth channel (upper right), and \thth channel (lower).
    The ratio also shows the signal scaled to the observed exclusion limit (as shown in Table~\ref{tab:hhxslimit}).
}
\label{fig:sob_channels}
\end{figure*}

\begin{figure*}[!htb]
\centering
\includegraphics[width=0.48\textwidth]{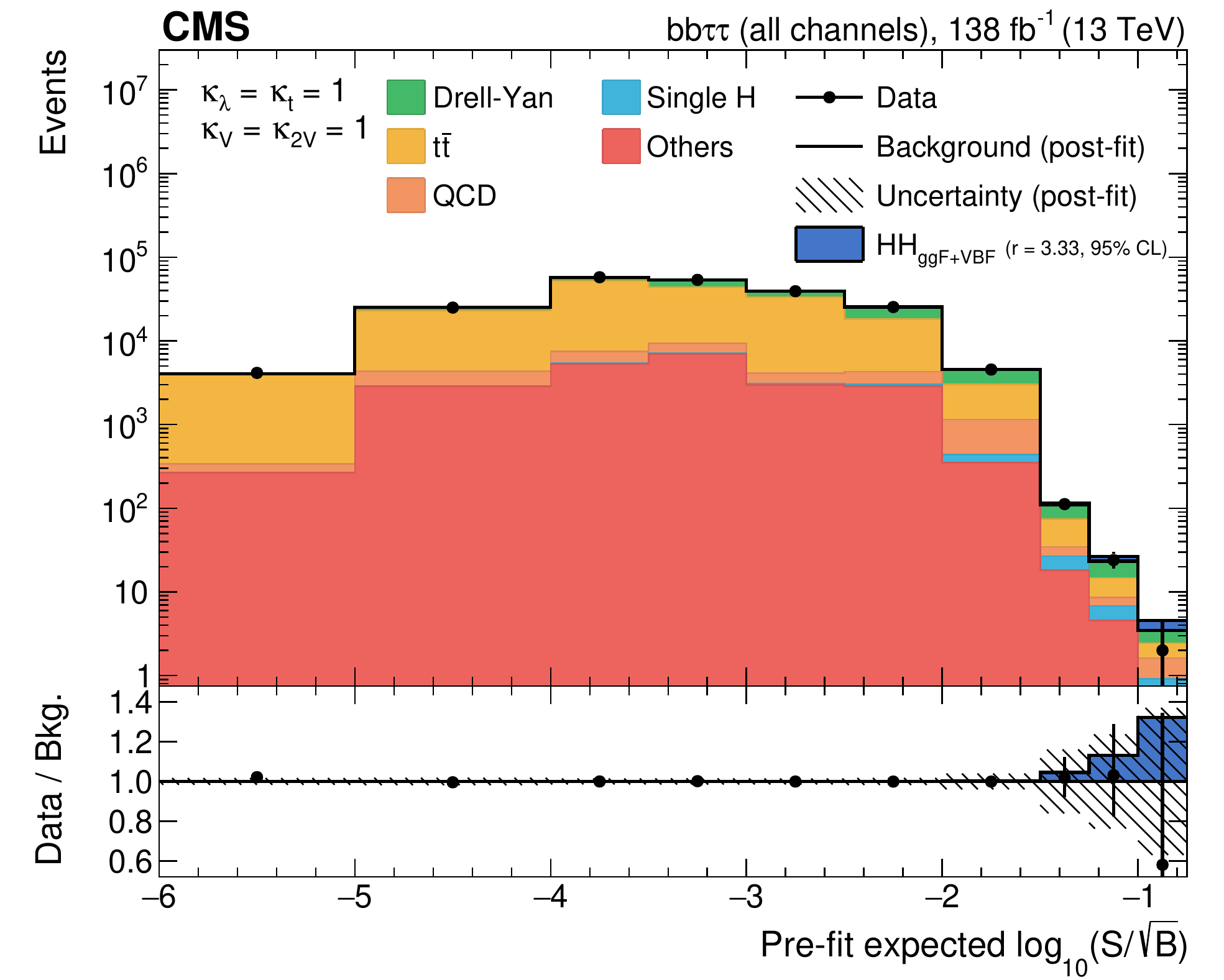}
\includegraphics[width=0.48\textwidth]{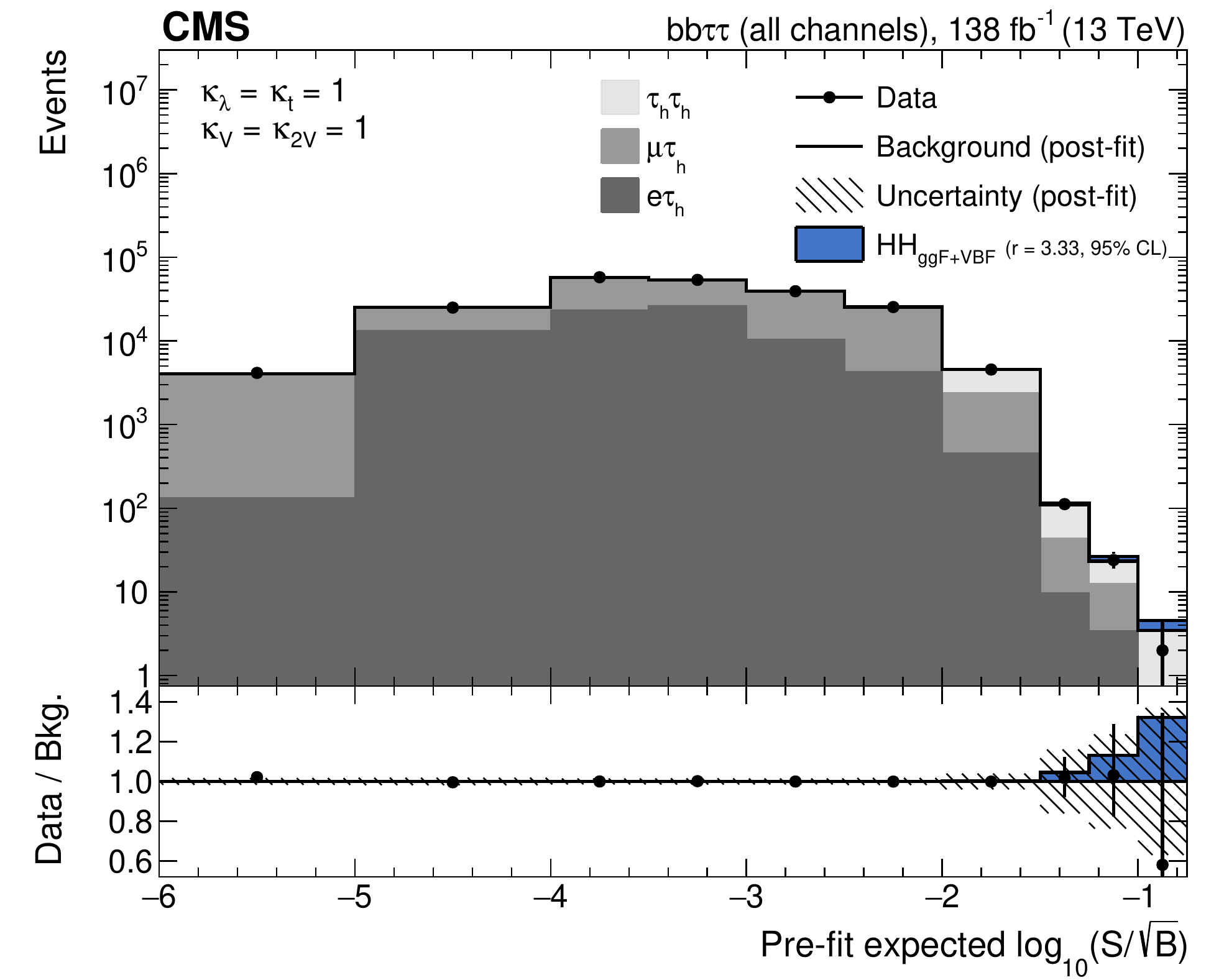}
\caption{
    Combination of bins of all postfit distributions, ordered according to the expected signal-to-square-root-background ratio,
    where the signal is the SM \HH signal expected in that bin and the background is the prefit background estimate in the same bin,
    separately for the background contribution split into physics processes (left), and split into the three considered final state channels (right).
    The ratio also shows the signal scaled to the observed exclusion limit (as shown in Table~\ref{tab:hhxslimit}).
}
\label{fig:sob_all}
\end{figure*}

As the total number of distributions is relatively high, we provide distributions that combine the content of bins with equal or similar
expected sensitivity in terms of a signal-to-noise ratio, as shown in Fig.~\ref{fig:sob_channels}, separately for the three channels
\eleth, \muth, and \thth, as well as in Fig.~\ref{fig:sob_all} for the combination of the full 2016--2018 data set.
In each plot, the bins of all input distributions are collected, then ordered according to their exected prefit signal-to-square-root-background ratio,
and eventually merged according to a dedicated binning rule. Hence, the right-most bins are most sensitive to the signal,
whereas the left-most bins are highly dominated by background contributions.

\begin{figure*}[!htb]
\centering
\includegraphics[width=0.49\textwidth]{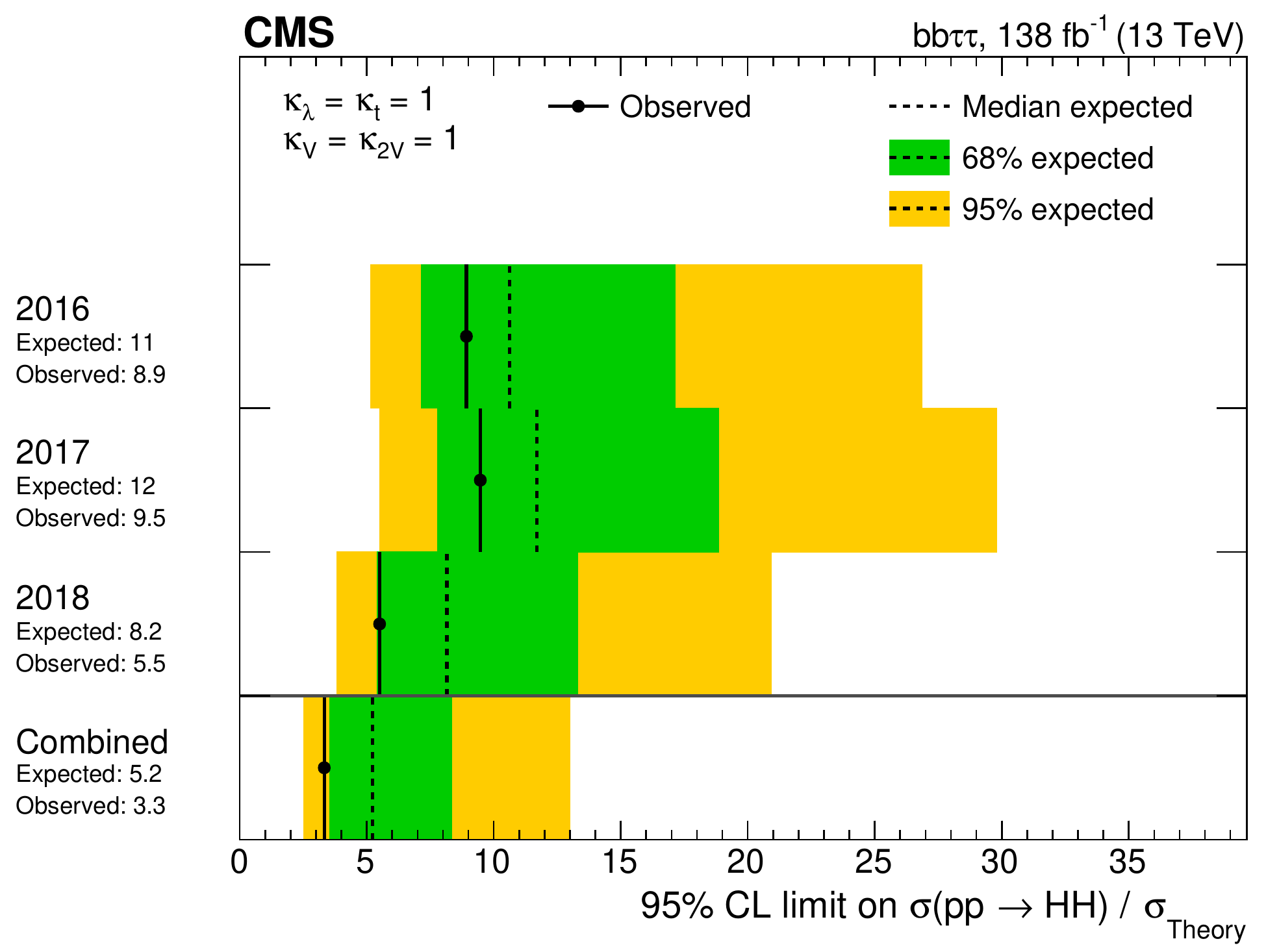}
\includegraphics[width=0.49\textwidth]{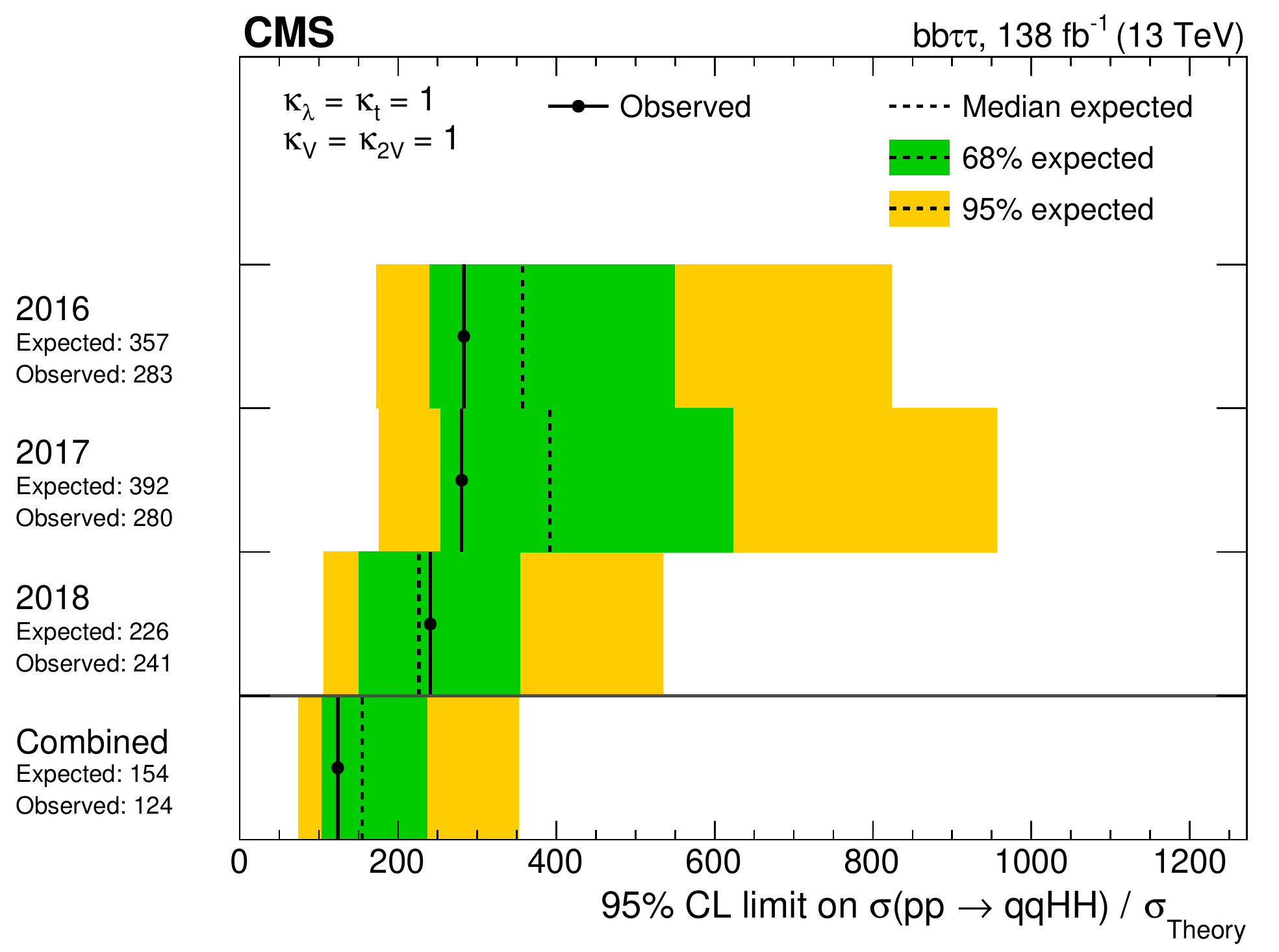}
\caption{
The expected and observed limits on the ratio of experimentally estimated \ggF{} plus VBF ($\sigma(\pp\to\HH)$, left) and VBF only ($\sigma(\pp\to\PQq\PQq\HH)$, right)
\HH production cross section and the expectation from the SM ($\sigma_{\text{Theory}}$) at 95\% \CL, separated into different years and combined for the full 2016--2018 data set.
}
\label{fig:hhxslimit}
\end{figure*}

The expected and observed limits for the \ggF{} plus VBF (inclusive) \HH production cross section at the SM point ($\kappa_{\lambda}=1$) and those for the
VBF only \HH production cross section ($\kappa_{2\PV}=1$) are listed in Tables~\ref{tab:hhxslimit} and~\ref{tab:vbfxslimit}, respectively.
A visual representation of these limits is provided in Fig.~\ref{fig:hhxslimit}.

\begin{table*}[!htb]
\centering
\small
\topcaption{Expected and observed upper limits at 95\% \CL for the SM point ($\kappa_{\lambda}=1$), where $\sigma_{\ggF+\text{VBF}}^{\text{SM}}=32.776\unit{fb}$ represents the sum of the \ggF{} plus VBF \HH cross sections.}
\renewcommand{\arraystretch}{1.2}
\begin{tabular}{lrrrr}
    Expected limit & \multicolumn{1}{c}{2016} & \multicolumn{1}{c}{2017} & \multicolumn{1}{c}{2018} & \multicolumn{1}{c}{Combined}  \\
    \hline
    $\sigma_{\ggF+\text{VBF}}(\pp\to\HH)/\sigma_{\ggF+\text{VBF}}^{\text{SM}}$   &  10.6 &  11.7 &   8.2 &   5.2 \\
    $\sigma_{\ggF+\text{VBF}}(\pp\to\HH)$ [\unit{fb}]                            & 324   & 356   & 249   & 159   \\
    $\sigma_{\ggF+\text{VBF}}(\pp\to\HH\to\bbtt)$ [\unit{fb}]                    &  23.6 &  26.0 &  18.2 &  11.6 \\ [\cmsTabSkip]
    Observed limit & \multicolumn{1}{c}{2016} & \multicolumn{1}{c}{2017} & \multicolumn{1}{c}{2018} & \multicolumn{1}{c}{Combined}  \\
    \hline
    $\sigma_{\ggF+\text{VBF}}(\pp\to\HH)/\sigma_{\ggF+\text{VBF}}^{\text{SM}}$   &   8.9 &  9.5  &   5.5 &   3.3 \\
    $\sigma_{\ggF+\text{VBF}}(\pp\to\HH)$ [\unit{fb}]                            & 272   & 291   & 169   & 102   \\
    $\sigma_{\ggF+\text{VBF}}(\pp\to\HH\to\bbtt)$ [\unit{fb}]                    &  19.6 &  21.2 &  12.4 &   7.5 \\
\end{tabular}
\label{tab:hhxslimit}
\end{table*}

\begin{table*}[!htb]
\centering
\small
\topcaption{Expected and observed upper limits at 95\% \CL for the SM point ($\kappa_{2\PV}=1$), where $\sigma_{\text{VBF}}^{\text{SM}}=1.726\unit{fb}$ represents the VBF only \HH cross section.}
\renewcommand{\arraystretch}{1.2}
\begin{tabular}{lrrrr}
    Expected limit & \multicolumn{1}{c}{2016} & \multicolumn{1}{c}{2017} & \multicolumn{1}{c}{2018} & \multicolumn{1}{c}{Combined}  \\
    \hline
    $\sigma_{\text{VBF}} (\pp\to\PQq\PQq\HH)/\sigma_{\text{VBF}}^{\text{SM}}$   & 357   & 392   & 226   & 154   \\
    $\sigma_{\text{VBF}} (\pp\to\PQq\PQq\HH)$ [\unit{fb}]                       & 616   & 676   & 391   & 266   \\
    $\sigma_{\text{VBF}} (\pp\to\PQq\PQq\HH\to\bbtt)$ [\unit{fb}]               &  45.0 &  49.3 &  28.5 &  19.4 \\ [\cmsTabSkip]
    Observed limit & \multicolumn{1}{c}{2016} & \multicolumn{1}{c}{2017} & \multicolumn{1}{c}{2018} & \multicolumn{1}{c}{Combined}  \\
    \hline
    $\sigma_{\text{VBF}} (\pp\to\PQq\PQq\HH)/\sigma_{\text{VBF}}^{\text{SM}}$   & 283   & 280   & 241   & 124   \\
    $\sigma_{\text{VBF}} (\pp\to\PQq\PQq\HH)$ [\unit{fb}]                       & 487   & 485   & 414   & 212   \\
    $\sigma_{\text{VBF}} (\pp\to\PQq\PQq\HH\to\bbtt)$ [\unit{fb}]               &  35.6 &  35.3 &  30.2 &  15.5 \\
\end{tabular}
\label{tab:vbfxslimit}
\end{table*}

\begin{figure*}[!htb]
\centering
\includegraphics[width=0.49\textwidth]{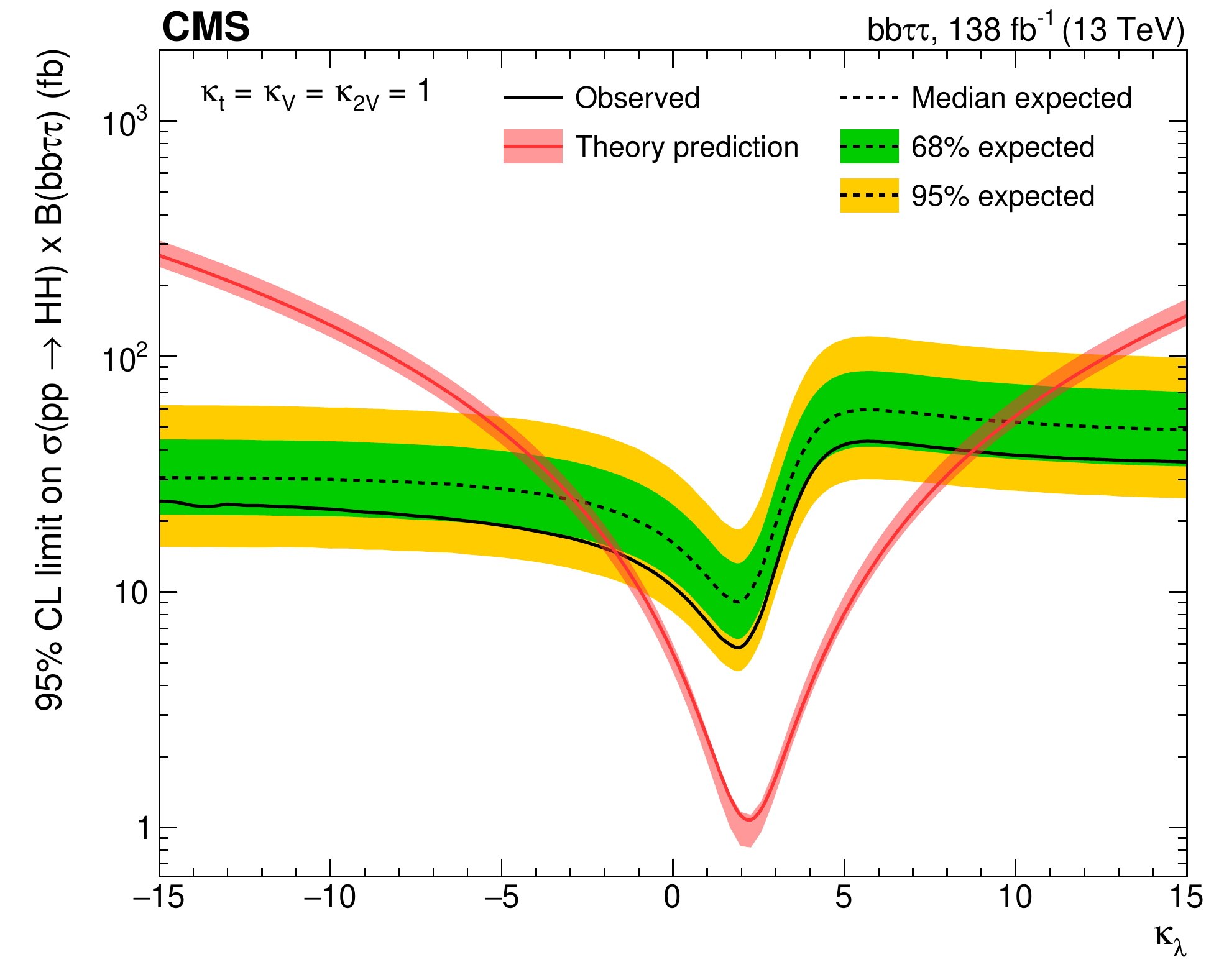}
\includegraphics[width=0.49\textwidth]{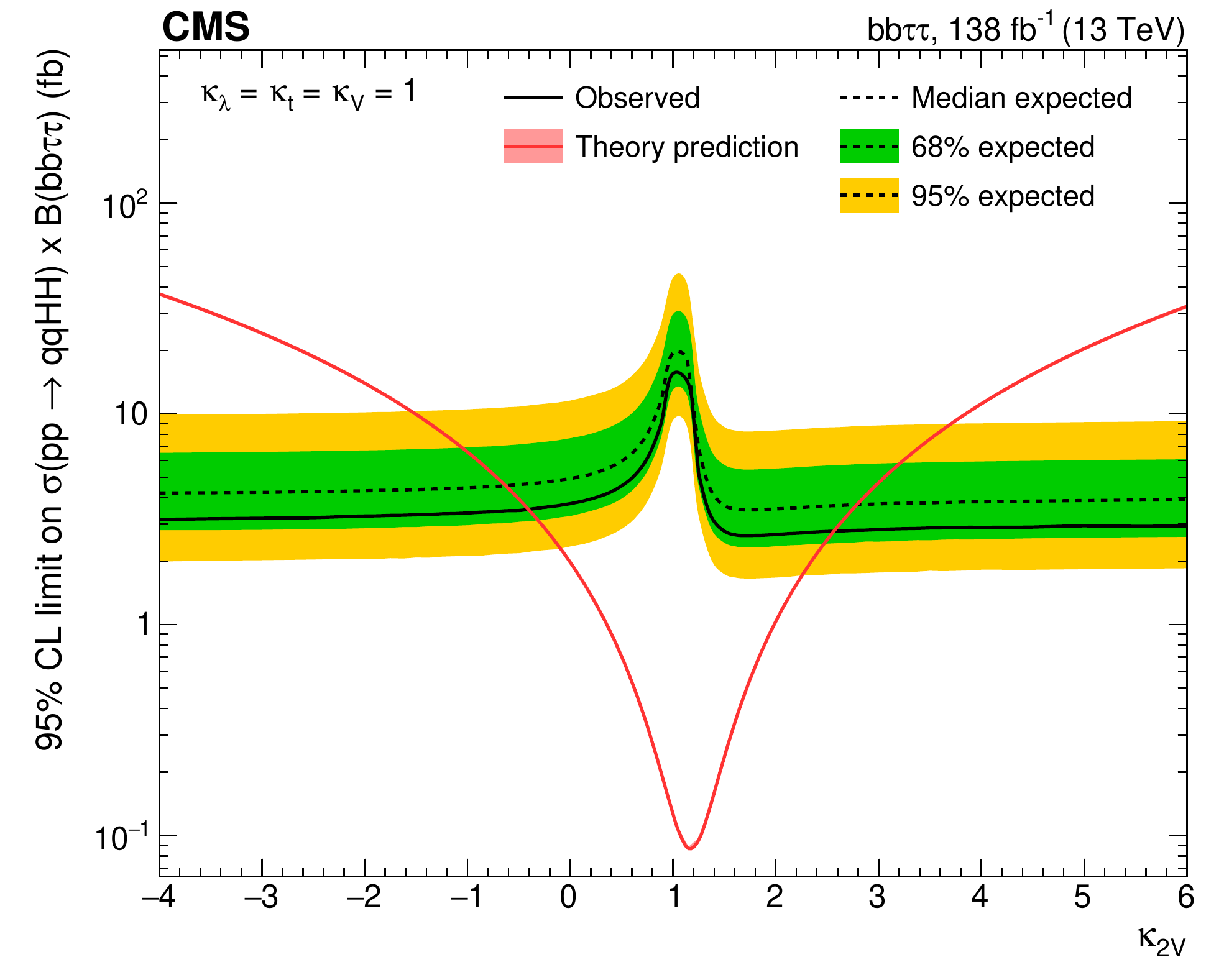}
\caption{
(left) Observed and expected upper limits at 95\% \CL as functions of $\kappa_{\lambda}$ on the \ggF{} plus VBF \HH cross section times the \bbtt branching fraction.
(right) Observed and expected upper limits at 95\% \CL as functions of $\kappa_{2\PV}$ on the VBF only \HH cross section times the \bbtt branching fraction.
In both cases all other couplings are set to their SM expectation.
The red solid line shows the theoretical prediction for the HH production cross section and its uncertainty (red shaded band).
}
\label{fig:hhlimit_scans}
\end{figure*}

From limits on the inclusive \HH cross section times the \bbtt branching fraction, computed for fixed values of $\kappa_{\lambda}$ following
the asymptotic \CLs prescription~\cite{CLS1,CLS2,Cowan:2010js}, we measure an observed (expected) 95\% \CL constraint on $\kappa_{\lambda}$ of
$-1.7<\kappa_{\lambda}<8.7$ ($-2.9<\kappa_{\lambda}<9.8$).
An observed (expected) 95\% \CL constraint on $\kappa_{2\PV}$ of $-0.4<\kappa_{2\PV}<2.6$ ($-0.6<\kappa_{2\PV}<2.8$) is instead deduced
from limits on the VBF only \HH cross section times the \bbtt branching fraction.
The corresponding scans are shown in Fig.~\ref{fig:hhlimit_scans}.

\begin{figure*}[!htb]
\centering
\includegraphics[width=0.49\textwidth]{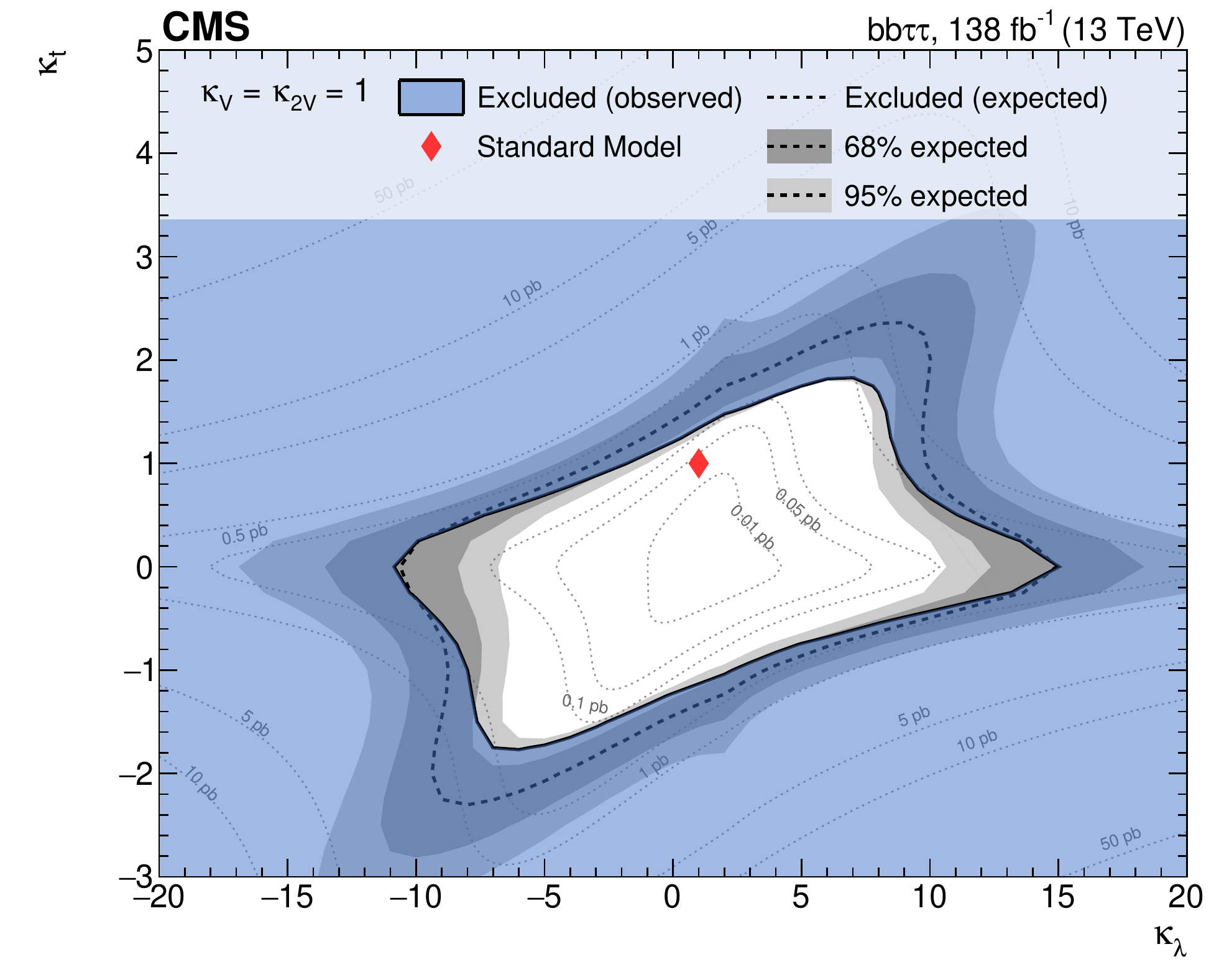}
\includegraphics[width=0.49\textwidth]{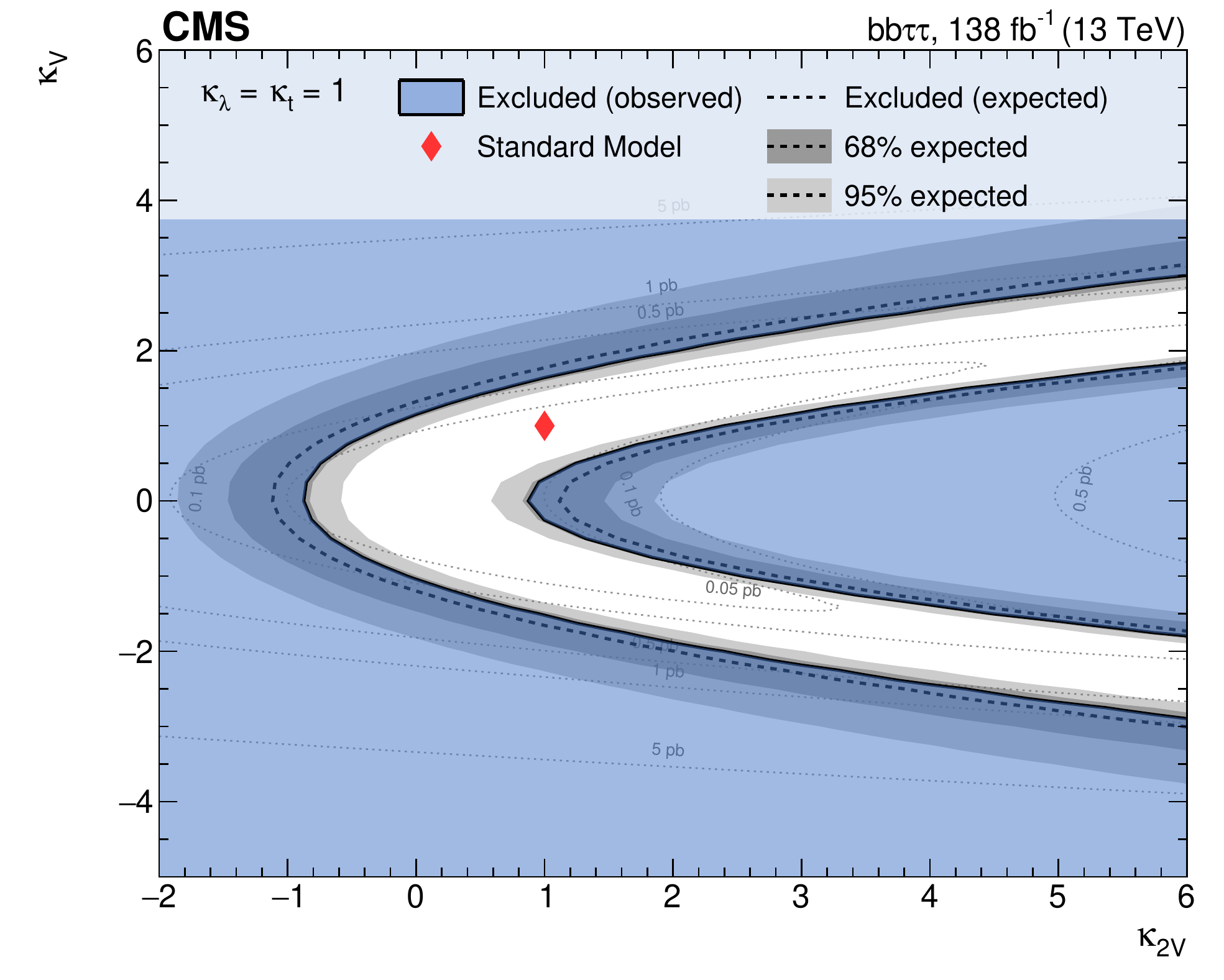}
\caption{(left) Two-dimensional exclusion regions as a function of the $\kappa_{\lambda}$ and $\kappa_t$ couplings for the full 2016--2018 combination, with both $\kappa_{2\PV}$
and $\kappa_{\PV}$ are fixed to unity. (right) Two-dimensional exclusion regions as a function of $\kappa_{2\PV}$ and $\kappa_{\PV}$,
with both $\kappa_{\lambda}$ and $\kappa_t$ are set to unity. Expected uncertainties on exclusion boundaries are inferred from uncertainty bands of the limit calculation,
and are denoted by dark and light-grey areas. The blue area marks parameter combinations that are observed to be excluded.
For visual guidance, theoretical cross section values are illustrated by thin, labeled contour lines with the SM prediction denoted by a red diamond.}
\label{fig:2Dexlusions}
\end{figure*}

The upper limits are used to set two-dimensional constraints on the coupling modifiers $\kappa_{\lambda}$, $\kappa_t$, $\kappa_{2\PV}$, and $\kappa_{\PV}$,
as shown in Fig.~\ref{fig:2Dexlusions}. A point in a two-dimensional parameter space is excluded when the upper limit on the
overall rate of \HH production at 95\% \CL is measured to be below the theoretical cross section, evaluated at the corresponding parameter values.

For both the \ggF{} and the VBF results here reported, the largest sensitivity is provided by the \thth channel, followed by the \muth channel, and lastly by
the \eleth one. The resolved categories, especially \textit{res2b}, dominate the sensitivity when measuring the constraint on $\kappa_{\lambda}$, while the combined
VBF multicategories, particularly \textit{classVBF}, provide the most stringent constraints on $\kappa_{2\PV}$.

\begin{figure*}[!htb]
\centering
\includegraphics[width=0.49\textwidth]{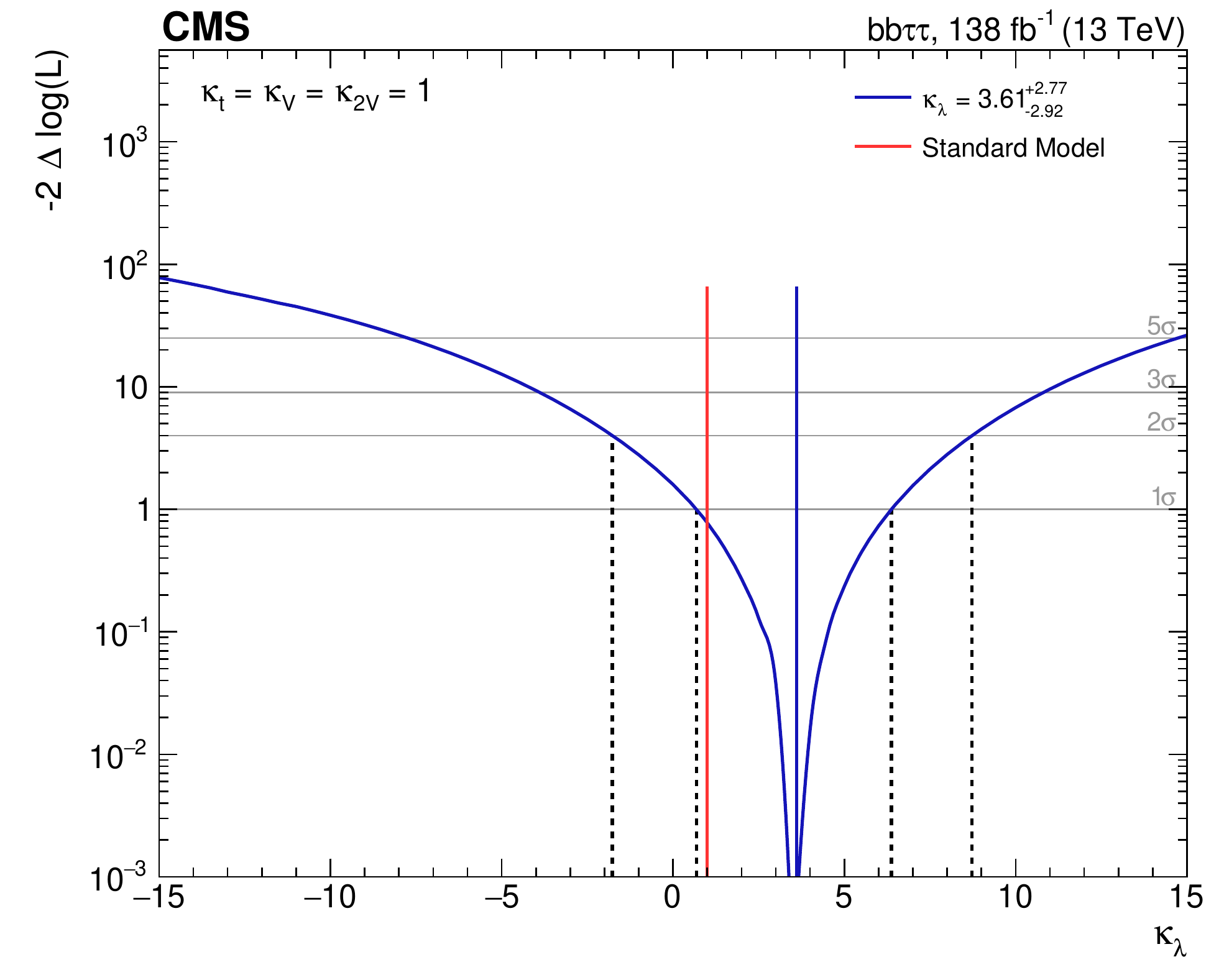}
\includegraphics[width=0.49\textwidth]{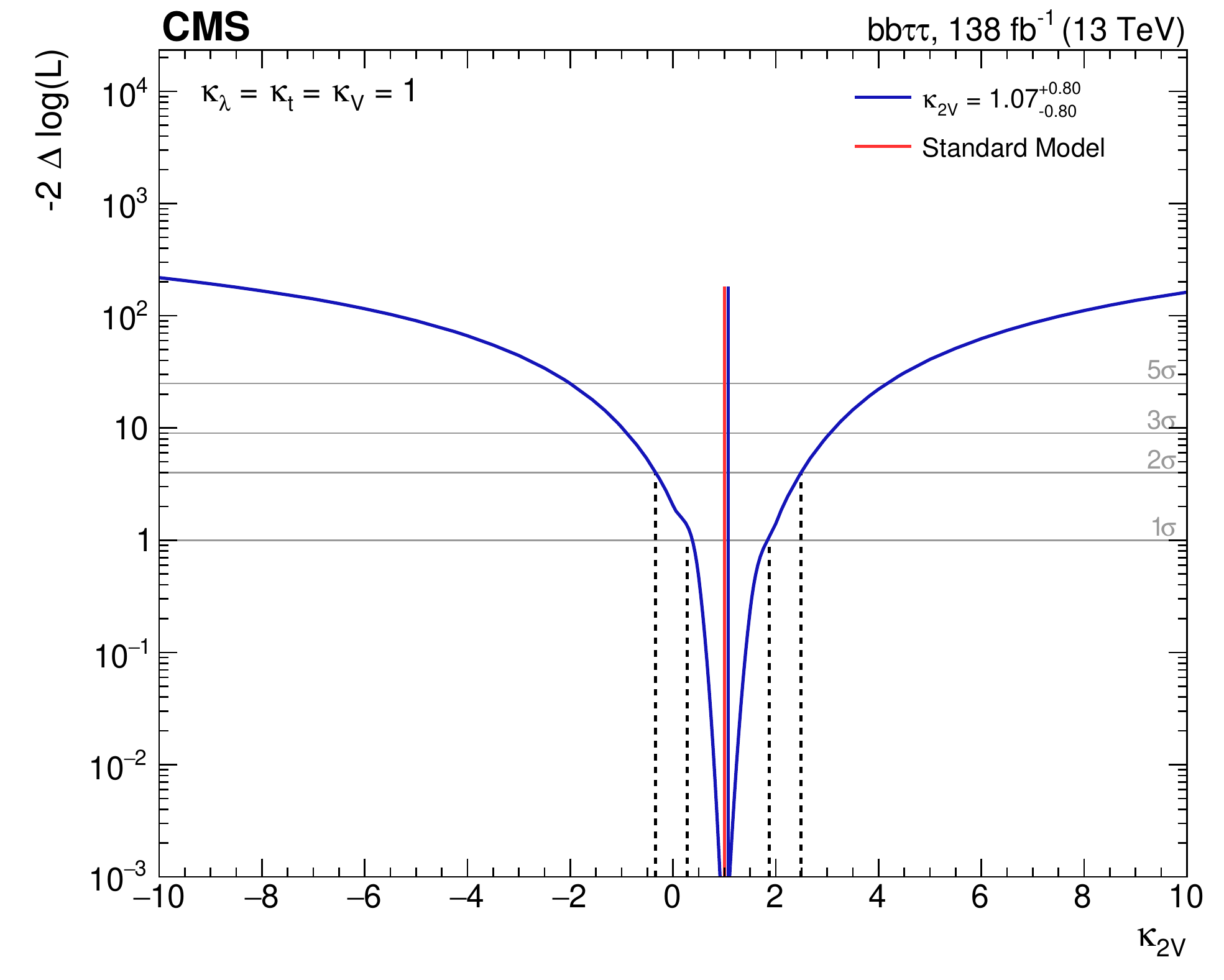}
\caption{Observed likelihood scan as a function of $\kappa_{\lambda}$ (left) and $\kappa_{2\PV}$ (right) for the full 2016--2018 combination.
The dashed lines show the intersection with threshold values one and four, corresponding to 68 and 95\% confidence intervals, respectively.}
\label{fig:1Dlikelihood}
\end{figure*}

Assuming that an \HH signal exists with the properties predicted by the SM, a scan of the likelihood as a function of the $\kappa_{\lambda}$ and $\kappa_{2\PV}$
coupling modifiers is performed, as shown in Fig.~\ref{fig:1Dlikelihood}.
This method consists in obtaining central best fit values and confidence intervals for the coupling modifiers directly from the profile likelihood~\cite{Cowan:2010js},
as opposed to using the \CLs method as in the results reported above.
The observed confidence interval on $\kappa_{\lambda}$ corresponds to $[+0.68,+6.37]$ at $68\%$ \CL and
to $[-1.77,+8.73]$ at $95\%$ \CL for a best fit value of $\kappa_{\lambda} = +3.61$. The observed confidence interval on $\kappa_{2\PV}$ corresponds
to $[+0.28,+1.88]$ at $68\%$ \CL and to $[-0.34,+2.49]$ at $95\%$ \CL for a best fit value of $\kappa_{2\PV}=+1.07$.

\ifthenelse{\boolean{cms@external}}{\clearpage}{}
\section{Summary}
A search for nonresonant Higgs boson pair (\HH) production via gluon-gluon fusion (\ggF) and vector boson fusion (VBF) processes in final states with two bottom quarks and two tau leptons was presented. The search uses the full 2016--2018 data set of proton-proton collisions at a center-of-mass energy of $\sqrt{s}=13\TeV$ recorded with the CMS detector at the LHC, corresponding to an integrated luminosity of 138\fbinv. The three decay modes of the $\PGt\PGt$ pair with the largest branching fraction have been selected, requiring one tau lepton to be always decaying hadronically and the other one either leptonically or hadronically. Upper limits at 95\% confidence level (\CL) on the inclusive \ggF{} plus VBF \HH production cross section are set, as well as on the VBF only \HH production cross section.

This analysis benefits from an improved trigger strategy as well as from a series of techniques developed especially for this search: among others, several neural networks to identify the \PQb jets from the \PH decay, to categorize the events, and to perform signal extraction. Moreover, this analysis builds up on the improvements made by the CMS Collaboration in the jet and tau lepton identification and reconstruction algorithms. All these techniques enable the achievement of particularly stringent results on the \HH production cross sections.

The observed (expected) 95\% \CL upper limit on \HH total production cross section corresponds to 3.3 (5.2) times the theoretical SM prediction. The observed (expected) 95\% \CL upper limit for the VBF only \HH SM cross section corresponds to 124 (154) times the theoretical SM prediction.

The observed (expected) 95\% \CL constraints on $\kappa_{\lambda}$ and $\kappa_{2\PV}$, derived from limits on the \HH production cross section times the \bbtt branching fraction, are found to be $-1.7<\kappa_{\lambda}<8.7$ ($-2.9<\kappa_{\lambda}<9.8$) and $-0.4<\kappa_{2\PV}<2.6$ ($-0.6<\kappa_{2\PV}<2.8$), respectively.

\begin{acknowledgments}
We congratulate our colleagues in the CERN accelerator departments for the excellent performance of the LHC and thank the technical and administrative staffs at CERN and at other CMS institutes for their contributions to the success of the CMS effort. In addition, we gratefully acknowledge the computing centers and personnel of the Worldwide LHC Computing Grid and other centers for delivering so effectively the computing infrastructure essential to our analyses. Finally, we acknowledge the enduring support for the construction and operation of the LHC, the CMS detector, and the supporting computing infrastructure provided by the following funding agencies: BMBWF and FWF (Austria); FNRS and FWO (Belgium); CNPq, CAPES, FAPERJ, FAPERGS, and FAPESP (Brazil); MES and BNSF (Bulgaria); CERN; CAS, MoST, and NSFC (China); MINCIENCIAS (Colombia); MSES and CSF (Croatia); RIF (Cyprus); SENESCYT (Ecuador); MoER, ERC PUT and ERDF (Estonia); Academy of Finland, MEC, and HIP (Finland); CEA and CNRS/IN2P3 (France); BMBF, DFG, and HGF (Germany); GSRI (Greece); NKFIH (Hungary); DAE and DST (India); IPM (Iran); SFI (Ireland); INFN (Italy); MSIP and NRF (Republic of Korea); MES (Latvia); LAS (Lithuania); MOE and UM (Malaysia); BUAP, CINVESTAV, CONACYT, LNS, SEP, and UASLP-FAI (Mexico); MOS (Montenegro); MBIE (New Zealand); PAEC (Pakistan); MES and NSC (Poland); FCT (Portugal); MESTD (Serbia); MCIN/AEI and PCTI (Spain); MOSTR (Sri Lanka); Swiss Funding Agencies (Switzerland); MST (Taipei); MHESI and NSTDA (Thailand); TUBITAK and TENMAK (Turkey); NASU (Ukraine); STFC (United Kingdom); DOE and NSF (USA).

\hyphenation{Rachada-pisek} Individuals have received support from the Marie-Curie program and the European Research Council and Horizon 2020 Grant, contract Nos.\ 675440, 724704, 752730, 758316, 765710, 824093, 884104, and COST Action CA16108 (European Union); the Leventis Foundation; the Alfred P.\ Sloan Foundation; the Alexander von Humboldt Foundation; the Belgian Federal Science Policy Office; the Fonds pour la Formation \`a la Recherche dans l'Industrie et dans l'Agriculture (FRIA-Belgium); the Agentschap voor Innovatie door Wetenschap en Technologie (IWT-Belgium); the F.R.S.-FNRS and FWO (Belgium) under the ``Excellence of Science -- EOS" -- be.h project n.\ 30820817; the Beijing Municipal Science \& Technology Commission, No. Z191100007219010; the Ministry of Education, Youth and Sports (MEYS) of the Czech Republic; the Hellenic Foundation for Research and Innovation (HFRI), Project Number 2288 (Greece); the Deutsche Forschungsgemeinschaft (DFG), under Germany's Excellence Strategy -- EXC 2121 ``Quantum Universe" -- 390833306, and under project number 400140256 - GRK2497; the Hungarian Academy of Sciences, the New National Excellence Program - \'UNKP, the NKFIH research grants K 124845, K 124850, K 128713, K 128786, K 129058, K 131991, K 133046, K 138136, K 143460, K 143477, 2020-2.2.1-ED-2021-00181, and TKP2021-NKTA-64 (Hungary); the Council of Science and Industrial Research, India; the Latvian Council of Science; the Ministry of Education and Science, project no. 2022/WK/14, and the National Science Center, contracts Opus 2021/41/B/ST2/01369 and 2021/43/B/ST2/01552 (Poland); the Funda\c{c}\~ao para a Ci\^encia e a Tecnologia, grant CEECIND/01334/2018 (Portugal); the National Priorities Research Program by Qatar National Research Fund; MCIN/AEI/10.13039/501100011033, ERDF ``a way of making Europe", and the Programa Estatal de Fomento de la Investigaci{\'o}n Cient{\'i}fica y T{\'e}cnica de Excelencia Mar\'{\i}a de Maeztu, grant MDM-2017-0765 and Programa Severo Ochoa del Principado de Asturias (Spain); the Chulalongkorn Academic into Its 2nd Century Project Advancement Project, and the National Science, Research and Innovation Fund via the Program Management Unit for Human Resources \& Institutional Development, Research and Innovation, grant B05F650021 (Thailand); the Kavli Foundation; the Nvidia Corporation; the SuperMicro Corporation; the Welch Foundation, contract C-1845; and the Weston Havens Foundation (USA).   
\end{acknowledgments}

\bibliography{auto_generated}

\providecommand{\href}[2]{#2}\begingroup\raggedright\begin{thebibliography}{10}%
\makeatletter
\providecommand{\hrefCMSnoop }[0]{\@secondoftwo}%
\makeatother
\providecommand{\doi}{\texttt{doi:}\begingroup \urlstyle{tt}\Url}

\bibitem{Aad:2012tfa}
\hrefCMSnoop {}{{ATLAS Collaboration}, ``Observation of a new particle in the
  search for the {Standard Model Higgs} boson with the {ATLAS} detector at the
  {LHC}'',} \textit{ Phys. Lett. B} \textbf{ 716} (2012) 1,
  \href{http://dx.doi.org/10.1016/j.physletb.2012.08.020}{\doi{10.1016/j.physletb.2012.08.020}},
\href{http://www.arXiv.org/abs/1207.7214}{\texttt{arXiv:1207.7214}}.

\bibitem{Chatrchyan:2012ufa}
\hrefCMSnoop {}{{CMS Collaboration}, ``Observation of a new boson at a mass of
  {125\GeV} with the {CMS} experiment at the {LHC}'',} \textit{ Phys. Lett. B}
  \textbf{ 716} (2012) 30,
  \href{http://dx.doi.org/10.1016/j.physletb.2012.08.021}{\doi{10.1016/j.physletb.2012.08.021}},
\href{http://www.arXiv.org/abs/1207.7235}{\texttt{arXiv:1207.7235}}.

\bibitem{Chatrchyan:2013lba}
\hrefCMSnoop {}{{CMS Collaboration}, ``Observation of a new boson with mass
  near {125\GeV} in proton-proton collisions at {$\sqrt{s}=7$ and 8\TeV}'',}
  \textit{ JHEP} \textbf{ 06} (2013) 081,
  \href{http://dx.doi.org/10.1007/JHEP06(2013)081}{\doi{10.1007/JHEP06(2013)081}},
\href{http://www.arXiv.org/abs/1303.4571}{\texttt{arXiv:1303.4571}}.

\bibitem{2020135425}
\hrefCMSnoop {}{{CMS Collaboration}, ``A measurement of the higgs boson mass in
  the diphoton decay channel'',} \textit{ Phys. Lett. B} \textbf{ 805} (2020)
  135425,
  \href{http://dx.doi.org/10.1016/j.physletb.2020.135425}{\doi{10.1016/j.physletb.2020.135425}},
  \href{http://www.arXiv.org/abs/2002.06398}{\texttt{arXiv:2002.06398}}.

\bibitem{Khachatryan:2016vau}
\hrefCMSnoop {}{{ATLAS, CMS} Collaboration, ``Measurements of the {Higgs} boson
  production and decay rates and constraints on its couplings from a combined
  {ATLAS} and {CMS} analysis of the {LHC} proton-proton collision data at
  {$\sqrt{s}=7$ and 8\TeV}'',} \textit{ JHEP} \textbf{ 08} (2016) 045,
  \href{http://dx.doi.org/10.1007/JHEP08(2016)045}{\doi{10.1007/JHEP08(2016)045}},
\href{http://www.arXiv.org/abs/1606.02266}{\texttt{arXiv:1606.02266}}.

\bibitem{Grazzini_2018}
M.~Grazzini\hrefCMSnoop {}{ {et~al.}, ``Higgs boson pair production at {NNLO}
  with top quark mass effects'',} \textit{ JHEP} \textbf{ 05} (2018) 59,
  \href{http://dx.doi.org/10.1007/jhep05(2018)059}{\doi{10.1007/jhep05(2018)059}},
\href{http://www.arXiv.org/abs/1803.02463}{\texttt{arXiv:1803.02463}}.

\bibitem{PhysRevD.103.056002}
J.~Baglio\hrefCMSnoop {}{ {et~al.}, ``{$gg\ensuremath{\rightarrow}HH$:}
  {Combined} uncertainties'',} \textit{ Phys. Rev. D} \textbf{ 103} (2021)
  056002,
  \href{http://dx.doi.org/10.1103/PhysRevD.103.056002}{\doi{10.1103/PhysRevD.103.056002}},
  \href{http://www.arXiv.org/abs/2008.11626v3}{\texttt{arXiv:2008.11626v3}}.

\bibitem{PhysRevD.98.114016}
\hrefCMSnoop {}{F.~A. Dreyer and A.~Karlberg, ``Vector-boson fusion {Higgs}
  pair production at {${\mathrm{N}}^{3}\mathrm{LO}$}'',} \textit{ Phys. Rev. D}
  \textbf{ 98} (2018) 114016,
  \href{http://dx.doi.org/10.1103/PhysRevD.98.114016}{\doi{10.1103/PhysRevD.98.114016}},
  \href{http://www.arXiv.org/abs/1811.07906}{\texttt{arXiv:1811.07906}}.

\bibitem{DiMicco:2692014}
B.~Di~Micco\hrefCMSnoop {}{ {et~al.}, ``Higgs boson pair production at
  colliders: status and perspectives'',} \textit{ Rev. Phys.} \textbf{ 5}
  (2020) 100045,
  \href{http://dx.doi.org/10.1016/j.revip.2020.100045}{\doi{10.1016/j.revip.2020.100045}},
  \href{http://www.arXiv.org/abs/1910.00012}{\texttt{arXiv:1910.00012}}.

\bibitem{CMS:2022prd}
\hrefCMSnoop {}{{CMS Collaboration}, ``Identification of hadronic tau lepton
  decays using a deep neural network'',} \textit{ JINST} \textbf{ 17} (2022)
  P07023,
  \href{http://dx.doi.org/10.1088/1748-0221/17/07/P07023}{\doi{10.1088/1748-0221/17/07/P07023}},
  \href{http://www.arXiv.org/abs/2201.08458}{\texttt{arXiv:2201.08458}}.

\bibitem{Sirunyan:2273798}
\hrefCMSnoop {}{{CMS Collaboration}, ``Search for {Higgs} boson pair production
  in events with two bottom quarks and two tau leptons in proton-proton
  collisions at {$\sqrt{s}=13\TeV$}'',} \textit{ Phys. Lett. B} \textbf{ 778}
  (2017) 101,
  \href{http://dx.doi.org/10.1016/j.physletb.2018.01.001}{\doi{10.1016/j.physletb.2018.01.001}},
  \href{http://www.arXiv.org/abs/1707.02909}{\texttt{arXiv:1707.02909}}.

\bibitem{PhysRevLett.121.191801}
\hrefCMSnoop {}{{ATLAS Collaboration}, ``Search for resonant and nonresonant
  {Higgs} boson pair production in the {$\PQb\PAQb\PGt^{+}\PGt^{-}$} decay
  channel in proton-proton collisions at {$\sqrt{s}=13\TeV$} with the {ATLAS}
  detector'',} \textit{ Phys. Rev. Lett.} \textbf{ 121} (2018) 191801,
  \href{http://dx.doi.org/10.1103/PhysRevLett.121.191801}{\doi{10.1103/PhysRevLett.121.191801}},
  \href{http://www.arXiv.org/abs/1808.00336}{\texttt{arXiv:1808.00336}}.

\bibitem{Bols_2020}
E.~Bols\hrefCMSnoop {}{ {et~al.}, ``Jet flavour classification using
  {DeepJet}'',} \textit{ JINST} \textbf{ 15} (2020) P12012,
  \href{http://dx.doi.org/10.1088/1748-0221/15/12/p12012}{\doi{10.1088/1748-0221/15/12/p12012}},
  \href{http://www.arXiv.org/abs/2008.10519}{\texttt{arXiv:2008.10519}}.

\bibitem{hepdata}
\hrefCMSnoop {}{``{HEPD}ata record for this analysis'',} 2022.
\newblock
  \href{http://dx.doi.org/10.17182/hepdata.127767}{\doi{10.17182/hepdata.127767}}.

\bibitem{Sirunyan:2020zal}
\hrefCMSnoop {}{{CMS Collaboration}, ``Performance of the {CMS Level-1} trigger
  in proton-proton collisions at {$\sqrt{s} = 13\TeV$}'',} \textit{ JINST}
  \textbf{ 15} (2020) P10017,
  \href{http://dx.doi.org/10.1088/1748-0221/15/10/P10017}{\doi{10.1088/1748-0221/15/10/P10017}},
  \href{http://www.arXiv.org/abs/2006.10165}{\texttt{arXiv:2006.10165}}.

\bibitem{Khachatryan:2016bia}
\hrefCMSnoop {}{{CMS Collaboration}, ``The {CMS} trigger system'',} \textit{
  JINST} \textbf{ 12} (2017) P01020,
  \href{http://dx.doi.org/10.1088/1748-0221/12/01/P01020}{\doi{10.1088/1748-0221/12/01/P01020}},
\href{http://www.arXiv.org/abs/1609.02366}{\texttt{arXiv:1609.02366}}.

\bibitem{Chatrchyan:2008zzk}
\hrefCMSnoop {}{{CMS Collaboration}, ``The {CMS} experiment at the {CERN}
  {LHC}'',} \textit{ JINST} \textbf{ 3} (2008) S08004,
  \href{http://dx.doi.org/10.1088/1748-0221/3/08/S08004}{\doi{10.1088/1748-0221/3/08/S08004}}.

\bibitem{CMS-LUM-17-003}
\hrefCMSnoop {}{{CMS Collaboration}, ``Precision luminosity measurement in
  proton-proton collisions at $\sqrt{s} =$ 13 {TeV} in 2015 and 2016 at
  {CMS}'',} \textit{ Eur. Phys. J. C} \textbf{ 81} (2021) 800,
  \href{http://dx.doi.org/10.1140/epjc/s10052-021-09538-2}{\doi{10.1140/epjc/s10052-021-09538-2}},
  \href{http://www.arXiv.org/abs/2104.01927}{\texttt{arXiv:2104.01927}}.

\bibitem{CMS-PAS-LUM-17-004}
\href {https://cds.cern.ch/record/2621960}{{CMS Collaboration}, ``{CMS}
  luminosity measurement for the 2017 data-taking period at {$\sqrt{s} =
  13\TeV$}'',} CMS Physics Analysis Summary CMS-PAS-LUM-17-004, CERN, 2017.

\bibitem{CMS-PAS-LUM-18-002}
\href {https://cds.cern.ch/record/2676164}{{CMS Collaboration}, ``{CMS}
  luminosity measurement for the 2018 data-taking period at {$\sqrt{s} =
  13\TeV$}'',} CMS Physics Analysis Summary CMS-PAS-LUM-18-002, CERN, 2019.

\bibitem{Alwall_2014}
J.~Alwall\hrefCMSnoop {}{ {et~al.}, ``The automated computation of tree-level
  and next-to-leading order differential cross sections, and their matching to
  parton shower simulations'',} \textit{ JHEP} \textbf{ 07} (2014) 79,
  \href{http://dx.doi.org/10.1007/jhep07(2014)079}{\doi{10.1007/jhep07(2014)079}},
  \href{http://www.arXiv.org/abs/1405.0301}{\texttt{arXiv:1405.0301}}.

\bibitem{Frixione_2007}
\hrefCMSnoop {}{S.~Frixione, P.~Nason, and C.~Oleari, ``Matching {NLO} {QCD}
  computations with parton shower simulations: the {POWHEG} method'',} \textit{
  JHEP} \textbf{ 11} (2007) 070,
  \href{http://dx.doi.org/10.1088/1126-6708/2007/11/070}{\doi{10.1088/1126-6708/2007/11/070}},
  \href{http://www.arXiv.org/abs/0709.2092}{\texttt{arXiv:0709.2092}}.

\bibitem{Re_2011}
\hrefCMSnoop {}{E.~Re, ``Single-top {Wt-channel} production matched with parton
  showers using the {\POWHEG} method'',} \textit{ Eur. Phys. J. C} \textbf{ 71}
  (2011) 1547,
  \href{http://dx.doi.org/10.1140/epjc/s10052-011-1547-z}{\doi{10.1140/epjc/s10052-011-1547-z}},
  \href{http://www.arXiv.org/abs/1009.2450}{\texttt{arXiv:1009.2450}}.

\bibitem{Campbell_2015}
\hrefCMSnoop {}{J.~M. Campbell, R.~K. Ellis, P.~Nason, and E.~Re, ``Top-pair
  production and decay at {NLO} matched with parton showers'',} \textit{ JHEP}
  \textbf{ 04} (2015) 114,
  \href{http://dx.doi.org/10.1007/jhep04(2015)114}{\doi{10.1007/jhep04(2015)114}},
  \href{http://www.arXiv.org/abs/1412.1828}{\texttt{arXiv:1412.1828}}.

\bibitem{Heinrich:2019bkc}
G.~Heinrich\hrefCMSnoop {}{ {et~al.}, ``{Probing the trilinear Higgs boson
  coupling in di-Higgs production at NLO QCD including parton shower
  effects}'',} \textit{ JHEP} \textbf{ 06} (2019) 66,
  \href{http://dx.doi.org/10.1007/JHEP06(2019)066}{\doi{10.1007/JHEP06(2019)066}},
  \href{http://www.arXiv.org/abs/1903.08137}{\texttt{arXiv:1903.08137}}.

\bibitem{Alwall_2007}
J.~Alwall\hrefCMSnoop {}{ {et~al.}, ``Comparative study of various algorithms
  for the merging of parton showers and matrix elements in hadronic
  collisions'',} \textit{ Eur. Phys. J. C} \textbf{ 53} (2007) 473,
  \href{http://dx.doi.org/10.1140/epjc/s10052-007-0490-5}{\doi{10.1140/epjc/s10052-007-0490-5}},
  \href{http://www.arXiv.org/abs/0706.2569}{\texttt{arXiv:0706.2569}}.

\bibitem{Frederix_2012}
\hrefCMSnoop {}{R.~Frederix and S.~Frixione, ``Merging meets matching in
  {MC@NLO}'',} \textit{ JHEP} \textbf{ 12} (2012) 61,
  \href{http://dx.doi.org/10.1007/jhep12(2012)061}{\doi{10.1007/jhep12(2012)061}},
  \href{http://www.arXiv.org/abs/1209.6215}{\texttt{arXiv:1209.6215}}.

\bibitem{Sj_strand_2015}
T.~Sj{\"o}strand\hrefCMSnoop {}{ {et~al.}, ``An introduction to {\PYTHIA}
  8.2'',} \textit{ Comp. Phys. Commun.} \textbf{ 191} (2015) 159,
  \href{http://dx.doi.org/10.1016/j.cpc.2015.01.024}{\doi{10.1016/j.cpc.2015.01.024}},
  \href{http://www.arXiv.org/abs/1410.3012}{\texttt{arXiv:1410.3012}}.

\bibitem{Khachatryan_2016}
\hrefCMSnoop {}{{CMS Collaboration}, ``Event generator tunes obtained from
  underlying event and multiparton scattering measurements'',} \textit{ Eur.
  Phys. J. C} \textbf{ 76} (2016) 155,
  \href{http://dx.doi.org/10.1140/epjc/s10052-016-3988-x}{\doi{10.1140/epjc/s10052-016-3988-x}},
  \href{http://www.arXiv.org/abs/1512.00815}{\texttt{arXiv:1512.00815}}.

\bibitem{Sirunyan_2020}
\hrefCMSnoop {}{{CMS Collaboration}, ``Extraction and validation of a new set
  of {CMS} {\PYTHIA}8 tunes from underlying-event measurements'',} \textit{
  Eur. Phys. J. C} \textbf{ 80} (2020) 4,
  \href{http://dx.doi.org/10.1140/epjc/s10052-019-7499-4}{\doi{10.1140/epjc/s10052-019-7499-4}},
  \href{http://www.arXiv.org/abs/1903.12179}{\texttt{arXiv:1903.12179}}.

\bibitem{NNPDF_3_0}
R.~D. Ball\hrefCMSnoop {}{ {et~al.}, ``Parton distributions for the {LHC} run
  {II}'',} \textit{ JHEP} \textbf{ 04} (2015) 40,
  \href{http://dx.doi.org/10.1007/JHEP04(2015)040}{\doi{10.1007/JHEP04(2015)040}},
  \href{http://www.arXiv.org/abs/1410.8849}{\texttt{arXiv:1410.8849}}.

\bibitem{NNPDF_3_1}
R.~D. Ball\hrefCMSnoop {}{ {et~al.}, ``Parton distributions from high-precision
  collider data'',} \textit{ Eur. Phys. J. C} \textbf{ 77} (2017) 663,
  \href{http://dx.doi.org/10.1140/epjc/s10052-017-5199-5}{\doi{10.1140/epjc/s10052-017-5199-5}},
  \href{http://www.arXiv.org/abs/1706.00428}{\texttt{arXiv:1706.00428}}.

\bibitem{AGOSTINELLI2003250}
\hrefCMSnoop {}{{GEANT4} Collaboration, ``{GEANT4}---a simulation toolkit'',}
  \textit{ Nucl. Instrum. Meth. A} \textbf{ 506} (2003) 250,
\href{http://dx.doi.org/10.1016/S0168-9002(03)01368-8}{\doi{10.1016/S0168-9002(03)01368-8}}.

\bibitem{CMS:2017yfk}
\hrefCMSnoop {}{{CMS Collaboration}, ``Particle-flow reconstruction and global
  event description with the {CMS} detector'',} \textit{ JINST} \textbf{ 12}
  (2017) P10003,
  \href{http://dx.doi.org/10.1088/1748-0221/12/10/P10003}{\doi{10.1088/1748-0221/12/10/P10003}},
\href{http://www.arXiv.org/abs/1706.04965}{\texttt{arXiv:1706.04965}}.

\bibitem{Sirunyan:2019kia}
\hrefCMSnoop {}{{CMS Collaboration}, ``Performance of missing transverse
  momentum reconstruction in proton-proton collisions at {$\sqrt{s} = 13\TeV$}
  using the {CMS} detector'',} \textit{ JINST} \textbf{ 14} (2019) P07004,
  \href{http://dx.doi.org/10.1088/1748-0221/14/07/P07004}{\doi{10.1088/1748-0221/14/07/P07004}},
\href{http://www.arXiv.org/abs/1903.06078}{\texttt{arXiv:1903.06078}}.

\bibitem{CMS-TDR-15-02}
\href {http://cds.cern.ch/record/2020886}{{CMS Collaboration}, ``Technical
  proposal for the {Phase-II} upgrade of the {Compact Muon Solenoid}'',} CMS
  Technical Proposal CERN-LHCC-2015-010, CMS-TDR-15-02, CERN, 2015.

\bibitem{CMS:2020uim}
\hrefCMSnoop {}{{CMS Collaboration}, ``Electron and photon reconstruction and
  identification with the {CMS} experiment at the {CERN} {LHC}'',} \textit{
  JINST} \textbf{ 16} (2021) P05014,
  \href{http://dx.doi.org/10.1088/1748-0221/16/05/P05014}{\doi{10.1088/1748-0221/16/05/P05014}},
  \href{http://www.arXiv.org/abs/2012.06888}{\texttt{arXiv:2012.06888}}.

\bibitem{CMS-DP-2020-021}
\href {https://cds.cern.ch/record/2717925}{{CMS Collaboration}, ``{ECAL} 2016
  refined calibration and {Run2} summary plots'',} CMS Detector Performance
  Summary CMS-DP-2020-021, CERN, 2020.

\bibitem{CMS:2018rym}
\hrefCMSnoop {}{{CMS Collaboration}, ``{Performance of the CMS muon detector
  and muon reconstruction with proton-proton collisions at $\sqrt{s}=$ 13
  TeV}'',} \textit{ JINST} \textbf{ 13} (2018) P06015,
  \href{http://dx.doi.org/10.1088/1748-0221/13/06/P06015}{\doi{10.1088/1748-0221/13/06/P06015}},
  \href{http://www.arXiv.org/abs/1804.04528}{\texttt{arXiv:1804.04528}}.

\bibitem{Chatrchyan:2012zz}
\hrefCMSnoop {}{{CMS Collaboration}, ``Performance of {$\PGt$}-lepton
  reconstruction and identification in {CMS}'',} \textit{ JINST} \textbf{ 7}
  (2012) P01001,
  \href{http://dx.doi.org/10.1088/1748-0221/7/01/P01001}{\doi{10.1088/1748-0221/7/01/P01001}},
\href{http://www.arXiv.org/abs/1109.6034}{\texttt{arXiv:1109.6034}}.

\bibitem{Khachatryan:2015dfa}
\hrefCMSnoop {}{{CMS Collaboration}, ``Reconstruction and identification of
  {$\PGt$} lepton decays to hadrons and {$\PGn_\PGt$} at {CMS}'',} \textit{
  JINST} \textbf{ 11} (2016) P01019,
  \href{http://dx.doi.org/10.1088/1748-0221/11/01/P01019}{\doi{10.1088/1748-0221/11/01/P01019}},
\href{http://www.arXiv.org/abs/1510.07488}{\texttt{arXiv:1510.07488}}.

\bibitem{Sirunyan:2637646}
\hrefCMSnoop {}{{CMS Collaboration}, ``Performance of reconstruction and
  identification of {$\PGt$} leptons decaying to hadrons and {$\PGn_\PGt$} in
  proton-proton collisions at {$\sqrt{s}=$ 13 TeV}'',} \textit{ JINST} \textbf{
  13} (2018) P10005,
  \href{http://dx.doi.org/10.1088/1748-0221/13/10/P10005}{\doi{10.1088/1748-0221/13/10/P10005}},
  \href{http://www.arXiv.org/abs/1809.02816}{\texttt{arXiv:1809.02816}}.

\bibitem{Cacciari:2008gp}
\hrefCMSnoop {}{M.~Cacciari, G.~P. Salam, and G.~Soyez, ``The anti-\kt jet
  clustering algorithm'',} \textit{ JHEP} \textbf{ 04} (2008) 063,
  \href{http://dx.doi.org/10.1088/1126-6708/2008/04/063}{\doi{10.1088/1126-6708/2008/04/063}},
  \href{http://www.arXiv.org/abs/0802.1189}{\texttt{arXiv:0802.1189}}.

\bibitem{Cacciari:2011ma}
\hrefCMSnoop {}{M.~Cacciari, G.~P. Salam, and G.~Soyez, ``{FastJet} user
  manual'',} \textit{ Eur. Phys. J. C} \textbf{ 72} (2012) 1896,
  \href{http://dx.doi.org/10.1140/epjc/s10052-012-1896-2}{\doi{10.1140/epjc/s10052-012-1896-2}},
\href{http://www.arXiv.org/abs/1111.6097}{\texttt{arXiv:1111.6097}}.

\bibitem{Dokshitzer:1997in}
\hrefCMSnoop {}{Y.~L. Dokshitzer, G.~D. Leder, S.~Moretti, and B.~R. Webber,
  ``Better jet clustering algorithms'',} \textit{ JHEP} \textbf{ 08} (1997)
  001,
  \href{http://dx.doi.org/10.1088/1126-6708/1997/08/001}{\doi{10.1088/1126-6708/1997/08/001}},
\href{http://www.arXiv.org/abs/hep-ph/9707323}{\texttt{arXiv:hep-ph/9707323}}.

\bibitem{Wobisch:1998wt}
\href {https://inspirehep.net/record/484872}{M.~Wobisch and T.~Wengler,
  ``Hadronization corrections to jet cross-sections in deep inelastic
  scattering'',} in \textit{ {Proceedings of the Workshop on Monte Carlo
  Generators for HERA Physics, Hamburg, Germany}}, p.~270.
\newblock 1998.
\newblock
\href{http://www.arXiv.org/abs/hep-ph/9907280}{\texttt{arXiv:hep-ph/9907280}}.
\newblock

\bibitem{Dasgupta:2013ihk}
\hrefCMSnoop {}{M.~Dasgupta, A.~Fregoso, S.~Marzani, and G.~P. Salam, ``Towards
  an understanding of jet substructure'',} \textit{ JHEP} \textbf{ 09} (2013)
  029,
  \href{http://dx.doi.org/10.1007/JHEP09(2013)029}{\doi{10.1007/JHEP09(2013)029}},
\href{http://www.arXiv.org/abs/1307.0007}{\texttt{arXiv:1307.0007}}.

\bibitem{Butterworth:2008iy}
\hrefCMSnoop {}{J.~M. Butterworth, A.~R. Davison, M.~Rubin, and G.~P. Salam,
  ``Jet substructure as a new {Higgs} search channel at the {LHC}'',} \textit{
  Phys. Rev. Lett.} \textbf{ 100} (2008) 242001,
  \href{http://dx.doi.org/10.1103/PhysRevLett.100.242001}{\doi{10.1103/PhysRevLett.100.242001}},
\href{http://www.arXiv.org/abs/0802.2470}{\texttt{arXiv:0802.2470}}.

\bibitem{Sirunyan:2020foa}
\hrefCMSnoop {}{{CMS Collaboration}, ``Pileup mitigation at {CMS} in
  {$\sqrt{s}=13\TeV$} data'',} \textit{ JINST} \textbf{ 15} (2020) P09018,
  \href{http://dx.doi.org/10.1088/1748-0221/15/09/p09018}{\doi{10.1088/1748-0221/15/09/p09018}},
  \href{http://www.arXiv.org/abs/2003.00503}{\texttt{arXiv:2003.00503}}.

\bibitem{Bertolini:2014bba}
\hrefCMSnoop {}{D.~Bertolini, P.~Harris, M.~Low, and N.~Tran, ``Pileup per
  particle identification'',} \textit{ JHEP} \textbf{ 10} (2014) 059,
  \href{http://dx.doi.org/10.1007/JHEP10(2014)059}{\doi{10.1007/JHEP10(2014)059}},
\href{http://www.arXiv.org/abs/1407.6013}{\texttt{arXiv:1407.6013}}.

\bibitem{Khachatryan:2016kdb}
\hrefCMSnoop {}{{CMS Collaboration}, ``Jet energy scale and resolution in the
  {CMS} experiment in proton-proton collisions at {$\sqrt{s}=8\TeV$}'',}
  \textit{ JINST} \textbf{ 12} (2017) P02014,
  \href{http://dx.doi.org/10.1088/1748-0221/12/02/P02014}{\doi{10.1088/1748-0221/12/02/P02014}},
\href{http://www.arXiv.org/abs/1607.03663}{\texttt{arXiv:1607.03663}}.

\bibitem{CMS-DP-2021-033}
\href {https://cds.cern.ch/record/2792322}{{CMS Collaboration}, ``Jet energy
  scale and resolution measurement with {Run-2} legacy data collected by {CMS}
  at {$\sqrt{s}=13\TeV$}'',} CMS Detector Performance Summary CMS-DP-2021-033,
  CERN, 2021.

\bibitem{1742-6596-513-2-022035}
\hrefCMSnoop {}{L.~Bianchini, J.~Conway, E.~K. Friis, and C.~Veelken,
  ``Reconstruction of the {Higgs} mass in {$\PH{}\to\PGt\PGt$} events by
  dynamical likelihood techniques'',} \textit{ J. Phys.: Conf. Ser.} \textbf{
  513} (2014) 022035,
  \href{http://dx.doi.org/10.1088/1742-6596/513/2/022035}{\doi{10.1088/1742-6596/513/2/022035}}.

\bibitem{higgsml_lumin}
\hrefCMSnoop {}{G.~C. Strong, ``On the impact of selected modern deep-learning
  techniques to the performance and celerity of classification models in an
  experimental high-energy physics use case'',} \textit{ Mach. Learn.: Sci.
  Technol.} (2020)
  \href{http://dx.doi.org/10.1088/2632-2153/ab983a}{\doi{10.1088/2632-2153/ab983a}},
  \href{http://www.arXiv.org/abs/2002.01427}{\texttt{arXiv:2002.01427}}.

\bibitem{SDR}
\hrefCMSnoop {}{S.~J. Hanson, ``A stochastic version of the delta rule'',}
  \textit{ Physica D} \textbf{ 42} (1990) 265,
  \href{http://dx.doi.org/10.1016/0167-2789(90)90081-Y}{\doi{10.1016/0167-2789(90)90081-Y}}.

\bibitem{dropout}
N.~Srivastava\href {http://jmlr.org/papers/v15/srivastava14a.html}{ {et~al.},
  ``Dropout: a simple way to prevent neural networks from overfitting'',}
  \textit{ JMLR} \textbf{ 15} (2014) 1929.

\bibitem{perceptron}
\hrefCMSnoop {}{F.~Rosenblatt, ``The perceptron: a probabilistic model for
  information storage and organization in the brain'',} \textit{ Psychological
  Review} \textbf{ 65} (1957) 386,
  \href{http://dx.doi.org/10.1037/h0042519}{\doi{10.1037/h0042519}}.

\bibitem{Linnainmaa_70}
\hrefCMSnoop {}{S.~Linnainmaa, ``Taylor expansion of the accumulated rounding
  error'',} \textit{ BIT Numerical Mathematics} \textbf{ 16} (1976) 146,
  \href{http://dx.doi.org/10.1007/BF01931367}{\doi{10.1007/BF01931367}}.

\bibitem{Werbos_81}
\hrefCMSnoop {}{P.~J. Werbos, ``Applications of advances in nonlinear
  sensitivity analysis'',} in \textit{ Proc. {10th IFIP Conf.}, 31/8--4/9,
  NYC}, p.~762.
\newblock 1981.

\bibitem{backprop}
\hrefCMSnoop {}{D.~E. {Rumelhart}, G.~E. {Hinton}, and R.~J. {Williams},
  ``Learning representations by back-propagating errors'',} \textit{ Nature}
  \textbf{ 323} (1986) 533,
  \href{http://dx.doi.org/10.1038/323533a0}{\doi{10.1038/323533a0}}.

\bibitem{CMS-HL-LHC}
\href {https://cds.cern.ch/record/2652549}{{CMS Collaboration}, ``Prospects for
  {HH} measurements at the {HL-LHC}'',} CMS Physics Analysis Summary
  CMS-PAS-FTR-18-019, CERN, Geneva, 2018.

\bibitem{Higgs-HL-LHC}
E.~M. Cepeda\hrefCMSnoop {}{ {et~al.}, ``Higgs physics at the {HL-LHC} and
  {HE-LHC}'',} \textit{ CERN Yellow Rep. Monogr.} \textbf{ 7} (2018) 221,
  \href{http://dx.doi.org/10.23731/CYRM-2019-007.221}{\doi{10.23731/CYRM-2019-007.221}},
  \href{http://www.arXiv.org/abs/1902.00134}{\texttt{arXiv:1902.00134}}.

\bibitem{YR4}
\hrefCMSnoop {}{D.~de~Florian {et~al.}, ``Handbook of {LHC Higgs} cross
  sections: 4. {Deciphering} the {Nature} of the {Higgs} sector'',} \textit{
  CERN Yellow Rep: Monogr.} \textbf{ 2} (2017) 21,
  \href{http://dx.doi.org/10.23731/CYRM-2017-002}{\doi{10.23731/CYRM-2017-002}},
  \href{http://www.arXiv.org/abs/1610.07922}{\texttt{arXiv:1610.07922}}.

\bibitem{dipoleShoweStudies}
\hrefCMSnoop {}{B.~Cabouat and T.~Sj{\"o}strand, ``Some dipole shower
  studies'',} \textit{ Eur. Phys. J. C} \textbf{ 78} (2018) 226,
  \href{http://dx.doi.org/10.1140/epjc/s10052-018-5645-z}{\doi{10.1140/epjc/s10052-018-5645-z}},
  \href{http://www.arXiv.org/abs/1710.00391v2}{\texttt{arXiv:1710.00391v2}}.

\bibitem{CLS1}
\hrefCMSnoop {}{T.~Junk, ``Confidence level computation for combining searches
  with small statistics'',} \textit{ Nucl. Instrum. Meth. A} \textbf{ 434}
  (1999) 435,
  \href{http://dx.doi.org/10.1016/S0168-9002(99)00498-2}{\doi{10.1016/S0168-9002(99)00498-2}},
\href{http://www.arXiv.org/abs/hep-ex/9902006}{\texttt{arXiv:hep-ex/9902006}}.

\bibitem{CLS2}
\hrefCMSnoop {}{A.~L. Read, ``Presentation of search results: the
  {CL$_{\text{s}}$} technique'',} \textit{ J. Phys. G} \textbf{ 28} (2002)
  2693,
\href{http://dx.doi.org/10.1088/0954-3899/28/10/313}{\doi{10.1088/0954-3899/28/10/313}}.

\bibitem{Cowan:2010js}
\hrefCMSnoop {}{G.~Cowan, K.~Cranmer, E.~Gross, and O.~Vitells, ``Asymptotic
  formulae for likelihood-based tests of new physics'',} \textit{ Eur. Phys. J.
  C} \textbf{ 71} (2011) 1554,
  \href{http://dx.doi.org/10.1140/epjc/s10052-011-1554-0}{\doi{10.1140/epjc/s10052-011-1554-0}},
  \href{http://www.arXiv.org/abs/1007.1727}{\texttt{arXiv:1007.1727}}.
[Erratum: \DOI{10.1140/epjc/s10052-013-2501-z}].

\bibitem{Mockus1991}
\hrefCMSnoop {}{J.~B. Mo{\v{c}}kus and L.~J. Mo{\v{c}}kus, ``Bayesian approach
  to global optimization and application to multiobjective and constrained
  problems'',} \textit{ Journal of Optimization Theory and Applications}
  \textbf{ 70} (1991) 157,
  \href{http://dx.doi.org/10.1007/BF00940509}{\doi{10.1007/BF00940509}},
  \href{http://www.arXiv.org/abs/2108.00002}{\texttt{arXiv:2108.00002}}.

\end{thebibliography}\endgroup
\cleardoublepage \appendix\section{The CMS Collaboration \label{app:collab}}\begin{sloppypar}\hyphenpenalty=5000\widowpenalty=500\clubpenalty=5000
\cmsinstitute{Yerevan Physics Institute, Yerevan, Armenia}
{\tolerance=6000
A.~Tumasyan\cmsAuthorMark{1}\cmsorcid{0009-0000-0684-6742}
\par}
\cmsinstitute{Institut f\"{u}r Hochenergiephysik, Vienna, Austria}
{\tolerance=6000
W.~Adam\cmsorcid{0000-0001-9099-4341}, J.W.~Andrejkovic, T.~Bergauer\cmsorcid{0000-0002-5786-0293}, S.~Chatterjee\cmsorcid{0000-0003-2660-0349}, K.~Damanakis\cmsorcid{0000-0001-5389-2872}, M.~Dragicevic\cmsorcid{0000-0003-1967-6783}, A.~Escalante~Del~Valle\cmsorcid{0000-0002-9702-6359}, P.S.~Hussain\cmsorcid{0000-0002-4825-5278}, M.~Jeitler\cmsAuthorMark{2}\cmsorcid{0000-0002-5141-9560}, N.~Krammer\cmsorcid{0000-0002-0548-0985}, L.~Lechner\cmsorcid{0000-0002-3065-1141}, D.~Liko\cmsorcid{0000-0002-3380-473X}, I.~Mikulec\cmsorcid{0000-0003-0385-2746}, P.~Paulitsch, F.M.~Pitters, J.~Schieck\cmsAuthorMark{2}\cmsorcid{0000-0002-1058-8093}, R.~Sch\"{o}fbeck\cmsorcid{0000-0002-2332-8784}, D.~Schwarz\cmsorcid{0000-0002-3821-7331}, S.~Templ\cmsorcid{0000-0003-3137-5692}, W.~Waltenberger\cmsorcid{0000-0002-6215-7228}, C.-E.~Wulz\cmsAuthorMark{2}\cmsorcid{0000-0001-9226-5812}
% ...
\par}
\vskip\cmsinstskip
\dag:~Deceased\\
$^{1}$Also at Yerevan State University, Yerevan, Armenia\\
$^{2}$Also at TU Wien, Vienna, Austria\\
% ...
\end{sloppypar}
\end{document}